\documentclass[secthm,seceqn,final]{elsart}
\journal{Journal of Differential Equations}
\usepackage{indentfirst}
\usepackage{amssymb,amsmath,mathrsfs}
\usepackage{cite}
\usepackage[bookmarks=true,pdfstartview=FitH,bookmarksnumbered=true,
bookmarksopen=open,bookmarksdepth=2,colorlinks=true,citecolor=red,linkcolor=blue
]{hyperref}

\allowdisplaybreaks
\newcommand{\norm}[1]{\|#1\|}
\newcommand{\abs}[1]{\left|#1\right|}
\newcommand{\set}[1]{\left\{#1\right\}}
\newcommand{\Real}{\mathbb R}
\newcommand{\NN}{\mathbb N}
\newcommand{\Z}{\mathbb Z}

\newcommand{\eps}{\varepsilon}

\newcommand{\Sz}{\mathscr{S}}
\newcommand{\F}{\mathscr{F}}
\newcommand{\C}{\mathcal{C}}

\newcommand{\ls}{\leqslant}
\newcommand{\gs}{\geqslant}

\newcommand{\supp}{\text{{\,\rm supp}\,}}
\newcommand{\dive}{\text{\,\rm div}}
\newcommand{\curl}{\text{\,\rm curl}}
\newcommand{\vu}{\mathbf{u}}
\newcommand{\vv}{\mathbf{v}}
\newcommand{\inc}{\mathbf{I}}
\newcommand{\comp}{\mathtt{c}}

\newcommand{\rhobar}{\bar{\rho}}

\newcommand{\te}[1][\rho]{{\tilde{#1}}}
\newcommand{\lam}{\Lambda}
\newcommand{\be}[1][s]{\dot{B}_{2,1}^{#1}}
\newcommand{\hbe}[2][s-1]{\tilde{B}_{2,1}^{#1,#2}}
\newcommand{\dk}[1][k]{\triangle_{#1}}
\newcommand{\propref}[1]{\hyperref[#1]{Proposition \ref{#1}}}
\newcommand{\thmref}[1]{\hyperref[#1]{Theorem \ref{#1}}}
\newcommand{\lemref}[1]{\hyperref[#1]{Lemma \ref{#1}}}
\newcommand{\secref}[1]{\hyperref[#1]{Section \ref{#1}}}
\newcommand{\en}{{E^{\frac{N}{2}}}}
\newcommand{\ent}{{E_T^{\frac{N}{2}}}}
\newcommand{\hn}{{\frac{N}{2}}}
\newcommand{\fr}{\mathscr{J}_n}

\renewenvironment{pf}[1][Proof]{{\par \noindent\textbf{#1.}}\;}{\hfill $\Box$\par}
\begin{document}

\begin{frontmatter}
\title{Global Existence for Compressible
Navier-Stokes-Poisson Equations in Three and Higher Dimensions}
\author[ch]{Chengchun Hao\corauthref{*}}
\ead{hcc@amss.ac.cn}
\corauth[*]{Corresponding author.}
\author[hl]{Hai-Liang Li}
\ead{hailiang.li.math@gmail.com}
\address[ch]{Institute of Mathematics,
   Academy of Mathematics \& Systems Science, CAS,\\
 Beijing 100190, P. R. China}
\address[hl]{Department of Mathematics,
Capital Normal University,\\
 Beijing 100037, P. R. China}



\begin{abstract}
The compressible Navier-Stokes-Poisson system is concerned in the
present paper, and the global existence and uniqueness of the strong
solution is shown in the framework of hybrid Besov spaces in three
and higher dimensions.
\end{abstract}
\end{frontmatter}

\section{Introduction}

In the present paper, we consider the Cauchy problem of the
following compressible Navier-Stokes-Poisson equations
\begin{align}\label{nsp}
\left\{\begin{aligned}
    &\rho_t+\dive(\rho \vu)=0,\\
    &(\rho \vu)_t+\dive (\rho \vu\otimes \vu)+\nabla P(\rho)
    =\rho\nabla\phi+\mu\Delta \vu+(\mu+\lambda)\nabla\dive \vu,\\
    &\Delta\phi=\rho-\rhobar,\\
    &(\rho,\vu)(0)=(\rho_0,\vu_0),
    \end{aligned}\right.
\end{align}
for $(t,x)\in [0,+\infty)\times \Real^N$, $N\gs 3$. $\rho$, $\vu$
and $\phi$ denote the electron density, electron velocity and the
electrostatic potential, respectively.
$P(\rho)=\frac{1}{2}\rho^\gamma$ is the pressure with $\gamma=2$.
$\mu,\,\lambda$ are the constant viscosity coefficients satisfying
$\mu>0$ and $2\mu+N\lambda\gs0$. The constant $\rhobar$ stands for
the density of positively charged background ions.
The Navier-Stokes-Poisson system is a simplified model (for
instance, the energy equation is not taken into granted) to describe
the dynamics of a charge transport where the compressible charged
fluid interacts with its own electric field against a charged ion
background \cite{Deg00}.
\par

Recently, many interesting researches have been devoted to
many topics of the compressible Navier-Stokes-Poisson (NSP) system. The
global existence of weak solutions of the compressible NSP system
subject to large initial data is shown \cite{Don03,ZhaT07}. The
quasi-neutral limits and related combining asymptotical limits
are proven \cite{DeJL07,DonM08,Ju-Li-Li,WangJ06}.  In the case that
the potential force representing the self-gravity in stellar gases,
the global existence of weak solutions and asymptotical behaviors
are also investigated recently, and the stability analysis for
compressible Navier-Stokes-Poisson and related systems is also
carried out, refer for instance to \cite{Du00,DFP04,DuZ05,KoS04} and
references therein.

The global existence of the classical solution is shown recently
\cite{Li-Mats-Zhang} in terms of the framework by Matsumura-Nishida.
In addition, the influence of the electric field is justified, which
affects the dissipation of the viscosity and the time-decay rate of
global solutions of IVP~\eqref{nsp} to the equilibrium state
$(\bar\rho,0)$, namely,
\begin{gather}
 c_{1}(1+t)^{-\frac34}\leq
 \|(\rho-\bar{\rho})(t)\|_{L^2(\Real^3)}
 \leq C(1+t)^{-\frac34},\label{xinde73}
\\
 c_{1}(1+t)^{-\frac14}
 \leq
 \|m(t)\|_{L^2(\Real^3)}
 \leq
 C(1+t)^{-\frac14},      \label{xinde7}
\end{gather}
where the decay rate of the momentum or the velocity is slower than the rate
$(1+t)^{-\frac34}$ for compressible Navier-Stokes equations. A
natural question follows then, that is, whether the similar
phenomena can be shown for global weak solutions or strong solutions
with lower regularity.

To this end, the first step is to show the global existence of
strong solutions in some Besov space with lower regularity. In this
paper, with the help of the classical Friedrichs' regularization
method, Littlewood-Paley analysis and hybrid Besov spaces, we are
able to construct the approximate solutions, obtain the a-priori
estimates in hybrid Besov spaces, and prove the global
existence of the unique  strong solution by the compactness
arguments as in \cite{CheM04,Dan01,Paicu}. Indeed, in terms of the
div-curl decomposition we can decompose the velocity vector field
into a vector field of the compressible part and a incompressible part.
Then, the original compressible system for the density and the velocity can
be decoupled into a system involving only the compressible system
for the ir-rotational (compressible) part of velocity vector field
and the electron density and the diffusion equation for the
divergence free (incompressible) part of velocity vector field as
used in~\cite{Dan01}. Thus, we can investigate the compressible
velocity field part and the incompressible velocity field part
separately to get the expected estimates in some hybrid Besov
spaces. As one can see later, however, the appearance of the electric
field leads to the rotational coupling effect and the loss of
regularity of density and the velocity vector field.

For simplicity, we only deal with the case $\gamma=2$, the arguments
used here can be applied to show the global existence for general
$\gamma> 1$. We have the main theorem as follows.

\begin{thm}\label{thm.main}
Let $N\gs 3$, $\gamma=2$, $\mu>0$ and $2\mu+N\lambda\gs0$. Assume
$\rho_0-\rhobar\in \hbe[\hn-\frac{5}{2}]{\hn}$ and
$\vu_0\in\hbe[\hn-\frac{3}{2}]{\hn-1}$.   Then, there exist two
positive constants $\alpha$ small enough and $M$ such that if
\begin{align*}
     \norm{\rho_0-\rhobar}_{\hbe[\hn-\frac{5}{2}]{\hn}}
     +\norm{\vu_0}_{\hbe[\hn-\frac{3}{2}]{\hn-1}}
     \ls\alpha,
\end{align*}
then \eqref{nsp} yields a unique global solution $(\rho,\vu,\phi)$
such that $(\rho-\rhobar,\vu,\phi)$ belongs to
\begin{align*}
    E:=\,&\C(\Real^+;\hbe[\hn-\frac{5}{2}]{\hn}
    \times(\hbe[\hn-\frac{3}{2}]{\hn-1})^N\times \hbe[\hn-\frac{1}{2}]{\hn+2})\\
    &\cap L^1(\Real^+;\hbe[\hn-\frac{1}{2}]{\hn}
    \times(\hbe[\hn+\frac{1}{2}]{\hn+1})^{N}
    \times \hbe[\hn+\frac{3}{2}]{\hn+2}),
\end{align*}
and satisfies
\begin{align*}
    \norm{(\rho-\rhobar,\vu,\phi)}_{E}
    \ls
    M(\norm{\rho_0-\rhobar}_{\hbe[\hn-\frac{5}{2}]{\hn}}
    +\norm{\vu_0}_{\hbe[\hn-\frac{3}{2}]{\hn-1}}),
\end{align*}x
where $M$ is independent of the initial data and the hybrid space $\hbe[s_1]{s_2}=\dot{B}_{2,1}^{s_{1}}\cap\dot{B}_{2,1}^{s_{2}}$ for $s_{1}\ls s_{2}$.
\end{thm}

The paper is organized as follows. We recall some Littlewood-Paley
theories for homogeneous Besov spaces and give the definitions and
some properties of hybrid Besov spaces in the second section. In
Sections \ref{sec.reform}-\ref{sec.linear}, we are dedicated into
reformulation of the system and proving a priori estimates for a
linearized system with convection terms. In Section \ref{sec.proof},
we prove the global existence and uniqueness of the solution.

\section{Littlewood-Paley decomposition and Besov spaces}

Let $\psi : \Real^N \to [0,1]$ be a radial smooth cut-off function
valued in $[0,1]$ such that
\begin{align*}
    \psi(\xi)=\left\{
    \begin{array}{ll}
    1, &\abs{\xi}\ls 3/4,\\
    \text{smooth}, &3/4<\abs{\xi}<4/3,\\
    0, &\abs{\xi}\gs 4/3.
    \end{array}
    \right.
\end{align*}
Let $\varphi(\xi)$ be the function
\begin{align*}
    \varphi(\xi):=\psi(\xi/2)-\psi(\xi).
\end{align*}
Thus, $\psi$ is supported in the ball $\set{\xi\in\Real^N:
\abs{\xi}\ls 4/3}$, and $\varphi$ is also a smooth cut-off function
valued in $[0,1]$ and supported in the annulus $\{\xi\in\Real^N:
3/4\ls\abs{\xi}\ls 8/3\}$.
 By construction, we have
\begin{align*}
    \sum_{k\in\Z}\varphi(2^{-k}\xi)=1, \quad \forall
    \xi\neq 0.
\end{align*}
One can define the dyadic blocks as follows. For $k\in\Z$, let
\begin{align*}
\dk f:=\F^{-1}\varphi(2^{-k}\xi)\F f.
\end{align*}
The formal decomposition
\begin{align}\label{lpd}
    f=\sum_{k\in\Z}\dk f
\end{align}
is called homogeneous Littlewood-Paley decomposition. Actually, this
decomposition works for just about any locally integrable function
which has some decay at infinity, and one usually has all the
convergence properties of the summation that one needs. Thus, the
r.h.s. of \eqref{lpd} does not necessarily converge in
$\Sz'(\Real^N)$. Even if it does, the equality is not always true in
$\Sz'(\Real^N)$. For instance, if $f\equiv 1$, then all the
projections $\dk f$ vanish. Nevertheless, \eqref{lpd} is true modulo
polynomials, in other words (cf.\cite{Dan05,Pee76}), if
$f\in\Sz'(\Real^N)$, then $\sum_{k\in\Z}\dk f$ converges modulo
$\mathscr{P}[\Real^N]$ and \eqref{lpd} holds in
$\Sz'(\Real^N)/\mathscr{P}[\Real^N]$.

\begin{defn}
Let $s\in\Real$, $1\ls p,\,q\ls \infty$. For $f\in\Sz'(\Real^N)$, we
write
\begin{align*}
    \norm{f}_{\be[s]}=\sum_{k\in\Z}
    2^{ks}\norm{\dk f}_{L^2}.
\end{align*}
\end{defn}

A difficulty comes from the choice of homogeneous spaces at this point.
Indeed, $\norm{\cdot}_{\be[s]}$ cannot be a norm on
$\{f\in\Sz'(\Real^N): \norm{f}_{\be[s]}<\infty\}$ because
$\norm{f}_{\be[s]}=0$ means that $f$ is a polynomial. This
enforces us to adopt the following definition for homogeneous Besov
spaces (cf. \cite{Dan01}).

\begin{defn}
Let $s\in\Real$ and $m=-[\hn+1-s]$. If $m<0$, then we define
$\be[s](\Real^N)$ as
\begin{align*}
    \be[s]=\Big\{f\in\Sz'(\Real^N):
    \norm{f}_{\be[s]}<\infty \text{ and } u=\sum_{k\in\Z}\dk f
    \text{ in } \Sz'(\Real^2)\Big\}.
\end{align*}
If $m\gs 0$, we denote by $\mathscr{P}_m$ the set of two variables
polynomials of degree less than or equal to $m$ and define
\begin{align*}
    \be[s]=\Big\{f\in\Sz'(\Real^N)/\mathscr{P}_m:
    \norm{f}_{\be[s]}<\infty \text{ and }u=\sum_{k\in\Z}\dk f
    \text{ in } \Sz'(\Real^N)/\mathscr{P}_m\Big\}.
\end{align*}
\end{defn}

For the composition of functions, we have the following estimates.

\begin{lem}[\mbox{\cite[Lemma 2.7]{Dan01}}]\label{lem.comp}
Let $s>0$ and $u\in \be[s]\cap L^\infty$. Then, it holds

\mbox{\rm (i) } Let $F\in W_{loc}^{[s]+2,\infty}(\Real^N)$ with
$F(0)=0$. Then $F(u)\in\be[s]$. Moreover, there exists a function of
one variable $C_0$ depending only on $s$ and $F$, and such that
\begin{align*}
    \norm{F(u)}_{\be[s]}\ls
    C_0(\norm{u}_{L^\infty})\norm{u}_{\be[s]}.
\end{align*}

\mbox{\rm (ii)} If $u,\, v\in\be[\hn]$, $(v-u)\in \be[s]$ for
$s\in(-\hn,\hn]$ and $G\in W_{loc}^{[\hn]+3,\infty}(\Real^N)$
satisfies $G'(0)=0$, then $G(v)-G(u)\in \be[s]$ and there exists a
function of two variables $C$ depending only on $s$, $N$ and $G$,
and such that
\begin{align*}
    \norm{G(v)-G(u)}_{\be[s]}\ls C(\norm{u}_{L^\infty},
    \norm{v}_{L^\infty})\left(\norm{u}_{\be[\hn]}+\norm{v}_{\be[\hn]}\right)
    \norm{v-u}_{\be[s]}.
\end{align*}
\end{lem}

We also need hybrid Besov spaces for which regularity assumptions
are different in low frequencies and high frequencies \cite{Dan01}.
We are going to recall the definition of these new spaces and some
of their main properties.

\begin{defn}
Let $s,\,t\in\Real$. We define
\begin{align*}
    \norm{f}_{\hbe[s]{t}}=\sum_{k\ls 0}2^{ks}\norm{\dk f}_{L^2}
    +\sum_{k>0}2^{kt}\norm{\dk f}_{L^2}.
\end{align*}
Let $m=-[\hn+1-s]$, we then define
\begin{align*}
    \hbe[s]{t}(\Real^N)
 =&\set{f\in\Sz'(\Real^N): \norm{f}_{\hbe[s]{t}}<\infty},
  \quad \text{if } m<0,\\
    \hbe[s]{t}(\Real^N)
 =&\set{f\in\Sz'(\Real^N)/\mathscr{P}_m: \norm{f}_{\hbe[s]{t}}<\infty},
   \quad \text{if } m \gs 0.
\end{align*}
\end{defn}

\begin{lem} We have the following inclusions for hybrid Besov spaces.

\mbox{\rm (i)} We have $\hbe[s]{s}=\be[s]$.

\mbox{\rm (ii)} If $s\ls t$ then $\hbe[s]{t}=\be[s]\cap\be[t]$.
Otherwise, $\hbe[s]{t}=\be[s]+\be[t]$.

\mbox{\rm (iii)} The space $\hbe[0]{s}$ coincides with the usual
inhomogeneous Besov space $B_{2,1}^s$.

\mbox{\rm (iv)} If $s_1\ls s_2$ and $t_1\gs t_2$, then
$\hbe[s_1]{t_1}\hookrightarrow \hbe[s_2]{t_2}$.
\end{lem}

Let us now recall some useful estimates for the product in hybrid
Besov spaces.

\begin{lem}[\mbox{\cite[Proposition 2.10]{Dan01}}]\label{lem.fgbesov}
Let $s_1,\, s_2>0$ and $f,\,g\in L^\infty\cap \hbe[s_1]{s_2}$. Then
$fg\in \hbe[s_1]{s_2}$ and
\begin{align*}
    \norm{fg}_{\hbe[s_1]{s_2}}\lesssim
    \norm{f}_{L^\infty}\norm{g}_{\hbe[s_1]{s_2}}
    +\norm{f}_{\hbe[s_1]{s_2}}\norm{g}_{L^\infty}.
\end{align*}
Let $s_1, s_2, t_1, t_2\ls \hn$ such that $\min(s_1+s_2, t_1+t_2)>0$,
$f\in \hbe[s_1]{t_1}$ and $g\in
\hbe[s_2]{t_2}$. Then $fg\in \hbe[s_1+s_2-1]{t_1+t_2-1}$ and
\begin{align*}
    \norm{fg}_{\hbe[s_1+s_2-\hn]{t_1+t_2-\hn}}\lesssim
    \norm{f}_{\hbe[s_1]{t_1}}\norm{g}_{\hbe[s_2]{t_2}}.
\end{align*}
\end{lem}

For $\alpha,\beta\in\Real$, let us define the following characteristic function on $\Z$:
\begin{align*}
     \tilde{\varphi}^{\alpha,\beta}(r)=\left\{
     \begin{array}{ll}
        \alpha,\quad &\text{if } r\ls 0,\\
        \beta, & \text{if } r\gs 1.
     \end{array}
     \right.
\end{align*}
Then, we can recall the following lemma.
\begin{lem}[\mbox{\cite[Lemma~6.2]{Dan01}}]\label{lem.innner}
    Let $F$ be an homogeneous smooth function of degree $m$. Suppose that $-N/2<s_1,t_1,s_2,t_2\ls 1+N/2$. The following two estimates hold:
    \begin{align*}
    &\abs{(F(D)\dk(\vv\cdot \nabla a),F(D)\dk a)}\\
    &\qquad\qquad\lesssim c_k2^{-k(\tilde{\varphi}^{s_1,s_2}(k)-m)}\norm{\vv}_{\be[\hn+1]} \norm{a}_{\hbe[s_1]{s_2}}\norm{F(D)\dk a}_{L^2},\\
    &\abs{(F(D)\dk(\vv\cdot\nabla a),\dk b)+(\dk(\vv\cdot\nabla b),F(D)\dk a)}\\
    &\qquad\qquad\lesssim c_k\norm{\vv}_{\be[\hn+1]}\times\big( 2^{-k(\tilde{\varphi}_{t_1,t_2}(k)-m)}\norm{F(D)\dk a}_{L^2}\norm{b}_{\hbe[t_1]{t_2}}\\
    &\qquad\qquad\qquad+2^{-k(\tilde{\varphi}^{s_1,s_2}(k)-m)}\norm{a}_{\hbe[s_1]{s_2}}
    \norm{\dk b}_{L^2}\big),
    \end{align*}
where $(\cdot,\cdot)$ denotes the $L^2$-inner product, the operator $F(D)$ is defined by $F(D)f:=\F^{-1} F(\xi)\F f$ and $\sum_{k\in\Z} c_k\ls 1$.
\end{lem}

\section{Reformulation of the Original System}\label{sec.reform}

Let $\te=\rho-\rhobar$. Then \eqref{nsp} can be rewritten as
\begin{align}\label{nsp1}
\left\{\begin{aligned}
    &\te_t+\vu\cdot\nabla \te +\rhobar\dive\vu=-\te\dive\vu,\\
    &\vu_t+\vu\cdot\nabla\vu-\rhobar^{-1}\mu\Delta\vu
    -\rhobar^{-1}(\mu+\lambda)\nabla\dive\vu+\nabla \te-\nabla\phi\\
    &\qquad\qquad\qquad\qquad=-\frac{ \te}{\rhobar(\te+\rhobar)}
    (\mu\Delta \vu+(\mu+\lambda)\nabla\dive \vu),\\
    &\Delta\phi=\te.
\end{aligned}\right.
\end{align}

Denote $\lam^s z:=\F^{-1}\abs{\xi}^s\F z$ for all $s\in\Real$. Let
$\comp=\lam^{-1}\dive \vu$ be the ``compressible part'' of the velocity
and $\inc=\lam^{-1}\curl \vu$ be the ``incompressible part''. Then,
we have
\begin{align*}
    \vu=-\lam^{-1}\nabla \comp-\lam^{-1}\dive \inc,
\end{align*}
since $\dive\dive\inc=0$. In fact,
\begin{align*}
    \dive\inc=\lam^{-1}\dive\curl\vu
    =\lam^{-1}(\Delta\vu-\nabla
    \dive\vu),
\end{align*}
which yields
\begin{align*}
    \dive\dive\inc=\lam^{-1}\dive(\Delta\vu-\nabla
    \dive\vu)
    =\lam^{-1}(\dive\Delta\vu-\Delta\dive\vu)=0.
\end{align*}
Moreover,
\begin{align*}
    \curl\dive\inc=\Delta\inc.
\end{align*}
The first equation in \eqref{nsp1} is changed into
\begin{align}\label{nsp2}
    \te_t+\vu\cdot\nabla \te+\rhobar\lam
    \comp=-\te \dive\vu.
\end{align}
For the 2nd equation in \eqref{nsp1}, applying $\lam^{-1}\dive$ and
$\lam^{-1}\curl$ to both sides, respectively, we get
\begin{align}\label{nsp3}
\left\{\begin{aligned}
   & \comp_t
    -\rhobar^{-1}(2\mu+\lambda)\Delta \comp
    -\lam \te-\lam^{-1}\te\\
    &\qquad\quad=-\lam^{-1}\dive\left\{\vu\cdot\nabla\vu
    +\frac{\te}{\rhobar (\te+\rhobar)}
    (\mu\Delta \vu+(\mu+\lambda)\nabla\dive \vu)\right\},\\
    &\inc_t-\rhobar^{-1}\mu\Delta\inc\\
    &\qquad\quad
     =-\lam^{-1}\curl\left\{
        \vu\cdot\nabla\vu
       +\frac{\te}{\rhobar (\te+\rhobar)}
        (\mu\Delta \vu+(\mu+\lambda)\nabla\dive \vu)
     \right\},
    \end{aligned}\right.
\end{align}
where we have used the fact
\begin{align*}
    \curl\nabla f=(\partial_j\partial_i f-\partial_i\partial_j
    f)_{ij}=(0)_{ij}=\mathbf{0}
\end{align*}
for any function $f$.

Because the first equation of \eqref{nsp3} involves $\lam^{-1}\te$,
we denote $h=\lam^{-1}\te$. Then, we have
\begin{align}\label{nsp4}
\left\{\begin{aligned}
    &h_t+\lam^{-1}(\vu\cdot\nabla\lam h)+\rhobar \comp=F,\\
    & \comp_t+\vu\cdot\nabla \comp-\rhobar^{-1}(2\mu+\lambda)\Delta \comp
    -\lam^2 h-h
    =G,\\
    &\inc_t-\rhobar^{-1}\mu\Delta\inc
    =H,\\
    &\vu=-\lam^{-1}\nabla \comp-\lam^{-1}\dive\inc,
    \end{aligned}\right.
\end{align}
where
\begin{align*}
    &F=-\lam^{-1}(\lam h\dive\vu),\quad
    G=\vu\cdot\nabla \comp-\lam^{-1}\dive J, \quad
    H=-\lam^{-1}\curl J,\\
    &J=\vu\cdot\nabla\vu+\frac{\te}{\rhobar (\te+\rhobar)}
    (\mu\Delta \vu+(\mu+\lambda)\nabla\dive \vu).
\end{align*}

The third equation is, up to nonlinear terms, a mere heat equation
on $\inc$. We therefore expect to get appropriate estimates for the
incompressible part of the velocity via the following lemma.

\begin{lem}\label{lem.heat}
Let $s\in\Real$, $r\in[1,+\infty]$, and $u$ solve
\begin{align*}
    \left\{
    \begin{aligned}
    &u_t-\rhobar^{-1}\mu\Delta u=f,\\
    &u(0)=u_0.
    \end{aligned}
    \right.
\end{align*}
Then there exists $C>0$ depending only on $N$, $\rhobar^{-1}\mu$ and $r$
such that, for all $0<T\ls+\infty$,
\begin{align*}
    \norm{u}_{L_T^r(\be[s+2/r])}\ls
    C\left(\norm{u_0}_{\be[s]}+\norm{f}_{L_T^1(\be[s])}\right).
\end{align*}
Moreover, $u\in \C([0,T];\be[s])$.
\end{lem}

For the first two equations, which is a linear coupling system, we
can use the following lemma.

\begin{prop}\label{prop.lin}
Let $(h,\comp)$ be a solution of
\begin{align}\label{lin}
\left\{\begin{aligned}
    &h_t+\lam^{-1}(\vv\cdot\nabla\lam h)+\rhobar \comp=F,\\
    & \comp_t+\vv\cdot\nabla \comp-\rhobar^{-1}(2\mu+\lambda)\Delta \comp
    -\lam^2 h-h
    =G,
    \end{aligned}\right.
\end{align}
on $[0,T)$, $(3-N)/2<s\ls (N+1)/2$ and $V(t)=\int_0^t
\norm{\vv(\tau)}_{\be[\frac{N}{2}+1]}d\tau$. The following estimate
holds on $[0,T)$:
\begin{align*}
    &\norm{h(t)}_{\hbe{s+\frac{3}{2}}}+\norm{\comp(t)}_{\hbe{s-\frac{1}{2}}}
    +\int_0^t(\norm{h(\tau)}_{\hbe[s+1]{s+\frac{3}{2}}}
    +\norm{\comp(\tau)}_{\hbe[s+1]{s+\frac{3}{2}}})d\tau\\
    \ls& Ce^{CV(t)}\Big(\norm{h(0)}_{\hbe{s+\frac{3}{2}}}
    +\norm{\comp(0)}_{\hbe{s-\frac{1}{2}}}\\
    &\qquad\qquad+\int_0^t e^{-CV(\tau)}
        \left(\norm{F(\tau)}_{\hbe{s+\frac{3}{2}}}
    +\norm{G(\tau)}_{\hbe{s-\frac{1}{2}}}
    \right)d\tau\Big),
\end{align*}
where $C$ depends only on $N$ and $s$.
\end{prop}

Let us define the functional space for $2-N/2<s\ls N/2+1$:
\begin{align}\label{space}
\begin{aligned}
    &E^s=\C(\Real^+;\hbe[s-\frac{3}{2}]{s+1}
    \times(\hbe[s-\frac{3}{2}]{s-1})^N)
    \cap L^1(\Real^+;(\hbe[s+\frac{1}{2}]{s+1})^{1+N}),\\
    &\norm{(h,\vu)}_{E^s}=\norm{h}_{L^\infty(\hbe[s-\frac{3}{2}]{s+1})}
    +\norm{\vu}_{L^\infty(\hbe[s-\frac{3}{2}]{s-1})}\\
    &\qquad\qquad\qquad+\norm{h}_{L^1(\hbe[s+\frac{1}{2}]{s+1})}
    +\norm{\vu}_{L^1(\hbe[s+\frac{1}{2}]{s+1})}.
\end{aligned}
\end{align}
When the time variable $t$ describes a finite length interval
$[0,T]$, we will denote by $E_T^s$ and $\norm{\cdot}_{E_T^s}$ the
corresponding spaces and norms.

\section{The estimates for the linear model}\label{sec.linear}

This section is devoted to the proof of \propref{prop.lin}. Let
$(h,\comp)$ be a solution of \eqref{lin} and denote
$\te[f]:=e^{-KV(t)}f$ for any function $f$. Then the system
\eqref{lin} can be transformed into the following form:
\begin{align}\label{lin0}
    \left\{\begin{aligned}
    &\te[h]_t+\lam^{-1}(\vv\cdot\nabla\lam \te[h])+\rhobar \te[\comp]
    =\te[F]-KV'(t)\te[h],\\
    & \te[\comp]_t+\vv\cdot\nabla \te[\comp]-\rhobar^{-1}(2\mu+\lambda)\Delta
    \te[\comp]
    -\lam^2 \te[h]-\te[h]=\te[G]-KV'(t)\te[\comp].
    \end{aligned}\right.
\end{align}

 Applying the operator $\dk$ to the system \eqref{lin0} and denoting
$f_k:=\dk f$, we have the following system
\begin{align}\label{lin1}
    \left\{\begin{aligned}
    &\partial_t\te[h]_k+\lam^{-1}(\vv\cdot\nabla\lam \te[h])_k
    +\rhobar \te[\comp]_k
    =\te[F]_k-KV'(t)\te[h]_k,\\
    &\partial_t \te[\comp]_k+(\vv\cdot\nabla \te[\comp])_k
    -\rhobar^{-1}(2\mu+\lambda)\Delta \te[\comp]_k
    -\lam^2 \te[h]_k-\te[h]_k  =\te[G]_k-KV'(t)\te[\comp]_k,
    \end{aligned}\right.
\end{align}

\subsection{Low frequencies $(k\ls 0)$}
Taking the $L^2$ scalar product of the first equation of
\eqref{lin1} with $\te[h]_k$, of the second equation with
$\te[\comp]_k$, we get the following two identities:
\begin{align}\label{lin2}
    \left\{\begin{aligned}
    &\frac{1}{2}\frac{d}{dt}\norm{\te[h]_k}_{L^2}^2
    +\rhobar (\te[\comp]_k,\te[h]_k)
    =(\te[F]_k,\te[h]_k)-KV'(t)\norm{\te[h]_k}_{L^2}^2\\
    &\qquad\qquad\qquad \qquad\qquad\qquad
    -(\lam^{-1}(\vv\cdot\nabla\lam \te[h])_k,\te[h]_k),\\
    &\frac{1}{2}\frac{d}{dt} \norm{\te[\comp]_k}_{L^2}^2
    +\rhobar^{-1}(2\mu+\lambda)\norm{\lam \te[\comp]_k}_{L^2}^2
    -(\lam \te[h]_k,\lam \te[\comp]_k)-(\te[h]_k,\te[\comp]_k)\\
    &\qquad\qquad\qquad =(\te[G]_k,\te[\comp]_k)-KV'(t)\norm{\te[\comp]_k}_{L^2}^2
    -((\vv\cdot\nabla \te[\comp])_k,\te[\comp]_k).
    \end{aligned}\right.
\end{align}
Now we want to get an equality involving $\lam \te[h]_k$. To achieve it,
we take $L^2$ scalar product of the first equation of \eqref{lin1}
with $\lam^2 \te[h]_k$, $\lam^4 \te[h]_k$ and $\lam^2 \te[\comp]_k$ and of the second
equation with $\lam^2 \te[h]_k$ and then sum the last two resulting
equalities, which yields, with the Plancherel theorem, that
\begin{align}\label{lin3}
   \left\{\begin{aligned}
   &\frac{1}{2}\frac{d}{dt}\norm{\lam \te[h]_k}_{L^2}^2+\rhobar(\lam
   \te[\comp]_k,\lam \te[h]_k)\\
   &\qquad=(\lam \te[F]_k,\lam \te[h]_k)-KV'(t)\norm{\lam\te[h]_k}_{L^2}^2
    -((\vv\cdot\nabla\lam \te[h])_k,\lam\te[h]_k),\\
   &\frac{d}{dt}(\lam^2 \te[h]_k, \te[\comp]_k)+\rhobar\norm{\lam \te[\comp]_k}_{L^2}^2
   +\rhobar^{-1}(2\mu+\lambda)(\lam^2 \te[h]_k,\lam^2 \te[\comp]_k)
   -\norm{\lam^2 \te[h]_k}_{L^2}^2   \\
   &\qquad-\norm{\lam \te[h]_k}_{L^2}^2
    =(\lam \te[F]_k,\lam \te[\comp]_k)-2KV'(t)(\lam^2 \te[h]_k,\te[\comp]_k)
    \\
    &\qquad\qquad-((\vv\cdot\nabla\lam \te[h])_k,\lam\te[\comp]_k)
    +(\lam G_k,\lam h_k)-(\lam(\vv\cdot\nabla\te[\comp])_k,\lam\te[h]_k).
   \end{aligned}\right.
\end{align}

 A linear combination
of \eqref{lin2} and \eqref{lin3} yields
\begin{align*}
    &\frac{1}{2}\frac{d}{dt}\Big[\frac{1}{\rhobar}\norm{\te[h]_k}_{L^2}^2
    +\frac{1}{\rhobar}\norm{\lam \te[h]_k}_{L^2}^2
    +\norm{\te[\comp]_k}_{L^2}^2-2K_1(\lam^2 \te[h]_k, \te[\comp]_k)\Big]
    +K_1\norm{\lam^2 \te[h]_k}_{L^2}^2\\
    &+K_1\norm{\lam \te[h]_k}_{L^2}^2
    +[\rhobar^{-1}(2\mu+\lambda)-\rhobar K_1]\norm{\lam \te[\comp]_k}_{L^2}^2
    -\rhobar^{-1}(2\mu+\lambda)K_1(\lam^3 \te[h]_k,\lam \te[\comp]_k)\\
   =&\frac{1}{\rhobar}\Big[(\te[F]_k,\te[h]_k)-KV'(t)\norm{\te[h]_k}_{L^2}^2
    -(\lam^{-1}(\vv\cdot\nabla\lam \te[h])_k,\te[h]_k)\Big]\\
   &+\frac{1}{\rhobar}\Big[(\lam \te[F]_k,\lam \te[h]_k)
    -KV'(t)\norm{\lam\te[h]_k}_{L^2}^2
    -((\vv\cdot\nabla\lam
    \te[h])_k,\lam\te[h]_k)\Big]\\
   &   +(\te[G]_k,\te[\comp]_k)-KV'(t)\norm{\te[\comp]_k}_{L^2}^2
    -((\vv\cdot\nabla \te[\comp])_k,\te[\comp]_k)\\
   &-K_1\Big[(\lam \te[F]_k,\lam \te[\comp]_k)+(\lam G_k,\lam h_k)
   -2KV'(t)(\lam^2 \te[h]_k,\te[\comp]_k)\\
    &\qquad\qquad\qquad-((\vv\cdot\nabla\lam \te[h])_k,
    \lam\te[\comp]_k)-(\lam(\vv\cdot\nabla\te[\comp])_k,\lam\te[h]_k)\Big].
\end{align*}
Noticing that $\norm{\lam \te[h]_k}_{L^2}\ls \frac{8}{3}2^k\norm{
\te[h]_k}_{L^2}\ls \frac{8}{3}\norm{\te[h]_k}_{L^2}$ for $k\ls 0$,
we have
\begin{align*}
    &\abs{(\lam^2 \te[h]_k, \te[\comp]_k)}\ls
    \frac{32M_1}{9}\norm{\lam \te[h]_k}_{L^2}^2
    +\frac{1}{2M_1}\norm{\te[\comp]_k}_{L^2}^2,\\
    &\abs{(\lam^3 \te[h]_k,\lam \te[\comp]_k)}\ls
    \frac{32M_2}{9}\norm{\lam^2 \te[h]_k}_{L^2}^2
    +\frac{1}{2M_2}\norm{\lam \te[\comp]_k}_{L^2}^2.
\end{align*}
Thus, we have to choose $K_1$, $M_1$ and $M_2$ satisfying
\begin{align*}
  &\frac{73}{64}-\frac{32}{9}\rhobar^{-1}(2\mu+\lambda)M_2>0,\quad
      K_1<M_1<\frac{\sqrt{3}}{8\sqrt{\rhobar}},\\  &2\mu+\lambda-\rhobar^{2}K_1-\frac{(2\mu+\lambda)K_1}{2M_2}>0.
\end{align*}
Hence, we can take
\begin{align*}
    M_1=\frac{1}{4\sqrt{\rhobar}}, \,
    M_2=\frac{5\rhobar}{16(2\mu+\lambda)}, \,
    K_1=\min\left(\frac{\rhobar(2\mu+\lambda)}{\rhobar^{3}+2(2\mu+\lambda)^2},
    \frac{1}{8\sqrt{\rhobar}}\right).
\end{align*}
Denote for $k\ls 0$
\begin{align*}
    \alpha_k^2:=\frac{1}{\rhobar}\norm{\te[h]_k}_{L^2}^2
    +\frac{1}{\rhobar}\norm{\lam \te[h]_k}_{L^2}^2
    +\norm{\te[\comp]_k}_{L^2}^2-2K_1(\lam^2 \te[h]_k, \te[\comp]_k).
\end{align*}
Then, there exist constants $c_3$ and $c_4$ such that
\begin{align*}
    c_1\alpha_k^2\ls\norm{\te[h]_k}_{L^2}^2
    +\norm{\lam \te[h]_k}_{L^2}^2
    +\norm{\te[\comp]_k}_{L^2}^2\ls
    c_2\alpha_k^2.
\end{align*}
Thus, there exists a constant $\hat{c}$ such that for $k\ls 0$
\begin{align}\label{lin6}
   \begin{aligned}
    &\frac{1}{2}\frac{d}{dt}\alpha_k^2+(\hat{c}2^{2k}+KV')\alpha_k^2\\
    \ls
     &\frac{1}{\rhobar}\Big[(\te[F]_k,\te[h]_k)
    -(\lam^{-1}(\vv\cdot\nabla\lam \te[h])_k,\te[h]_k)\Big]\\
    &+\frac{1}{\rhobar}\Big[(\lam \te[F]_k,\lam \te[h]_k)
    -((\vv\cdot\nabla\lam
    \te[h])_k,\lam\te[h]_k)\Big]\\
   &   +(\te[G]_k,\te[\comp]_k)
    -((\vv\cdot\nabla \te[\comp])_k,\te[\comp]_k)
    -K_1\Big[(\lam \te[F]_k,\lam \te[\comp]_k)+(\lam G_k,\lam h_k)   \\
    &\qquad\qquad\qquad-((\vv\cdot\nabla\lam \te[h])_k,
    \lam\te[\comp]_k)-(\lam(\vv\cdot\nabla\te[\comp])_k,\lam\te[h]_k)\Big].
   \end{aligned}
\end{align}

\subsection{High frequencies $(k>0)$.}

Taking the $L^2$ scalar product of the first equation of
\eqref{lin1} with $\lam\te[h]_k$, of the second equation with
$\lam\te[\comp]_k$, we get the following two identities:
\begin{align}\label{lin21}
    \left\{\begin{aligned}
    &\frac{1}{2}\frac{d}{dt}\norm{\lam^{\frac{1}{2}}\te[h]_k}_{L^2}^2
    +\rhobar (\te[\comp]_k,\lam\te[h]_k)\\
    &\qquad\qquad=(\te[F]_k,\lam\te[h]_k)
     -KV'(t)\norm{\lam^{\frac{1}{2}}\te[h]_k}_{L^2}^2
    -((\vv\cdot\nabla\lam \te[h])_k,\te[h]_k),\\
    &\frac{1}{2}\frac{d}{dt} \norm{\lam^{\frac{1}{2}}\te[\comp]_k}_{L^2}^2
    +\rhobar^{-1}(2\mu+\lambda)\norm{\lam^{\frac{3}{2}} \te[\comp]_k}_{L^2}^2
    -(\lam^2 \te[h]_k,\lam \te[\comp]_k)-(\lam\te[h]_k,\te[\comp]_k)\\
    &\qquad\qquad =(\lam^{\frac{1}{2}}\te[G]_k,\lam^{\frac{1}{2}}\te[\comp]_k)
    -KV'(t)\norm{\lam^{\frac{1}{2}}\te[\comp]_k}_{L^2}^2
    -((\vv\cdot\nabla \te[\comp])_k,\lam\te[\comp]_k).
    \end{aligned}\right.
\end{align}
Now we want to get an equality involving $\lam^3 \te[h]_k$. To achieve it,
we take $L^2$ scalar product of the first equation of \eqref{lin1}
with $\lam^3 \te[h]_k$, $\lam^5 \te[h]_k$ and $\lam^3 \te[\comp]_k$ and of the second
equation with $\lam^3 \te[h]_k$ and then sum the last two resulting
equalities, which yields, with the Plancherel theorem, that
\begin{align}\label{lin31}
   \left\{\begin{aligned}
   &\frac{1}{2}\frac{d}{dt}\norm{\lam^{\frac{3}{2}} \te[h]_k}_{L^2}^2+\rhobar(\lam
   \te[\comp]_k,\lam^2 \te[h]_k)\\
   &\qquad=(\lam \te[F]_k,\lam^2 \te[h]_k)-KV'(t)\norm{\lam^{\frac{3}{2}}\te[h]_k}_{L^2}^2
    -((\vv\cdot\nabla\lam \te[h])_k,\lam^2\te[h]_k),\\
   &\frac{1}{2}\frac{d}{dt}\norm{\lam^{\frac{5}{2}} \te[h]_k}_{L^2}^2+\rhobar(\lam^2
   \te[\comp]_k,\lam^3 \te[h]_k)\\
   &\qquad=(\lam^2 \te[F]_k,\lam^3 \te[h]_k)-KV'(t)\norm{\lam^{\frac{5}{2}}\te[h]_k}_{L^2}^2
    -(\lam(\vv\cdot\nabla\lam \te[h])_k,\lam^3\te[h]_k),\\
   &\frac{d}{dt}(\lam^3 \te[h]_k, \te[\comp]_k)+\rhobar\norm{\lam^{\frac{3}{2}} \te[\comp]_k}_{L^2}^2
   +\rhobar^{-1}(2\mu+\lambda)(\lam^3 \te[h]_k,\lam^2 \te[\comp]_k)\\
   &\qquad\qquad-\norm{\lam^{\frac{5}{2}} \te[h]_k}_{L^2}^2
   -\norm{\lam^{\frac{3}{2}} \te[h]_k}_{L^2}^2\\
   &\qquad=(\lam^2 \te[F]_k,\lam \te[\comp]_k)-2KV'(t)(\lam^3 \te[h]_k,\te[\comp]_k)
    -((\vv\cdot\nabla\lam \te[h])_k,\lam^2\te[\comp]_k)\\
    &\qquad\qquad+(\lam \te[G]_k,\lam^2 \te[h]_k)
     -(\lam(\vv\cdot\nabla\te[\comp])_k,\lam^2\te[h]_k).
   \end{aligned}\right.
\end{align}

 A linear combination of
\eqref{lin21} and \eqref{lin31} yields
\begin{align*}
    &\frac{1}{2}\frac{d}{dt}\Big[
    \frac{1}{\rhobar}\norm{\lam^{\frac{1}{2}}\te[h]_k}_{L^2}^2
    +\frac{1}{\rhobar}\norm{\lam^{\frac{3}{2}}\te[h]_k}_{L^2}^2
    +\rhobar^{-2}(2\mu+\lambda)K_2\norm{\lam^{\frac{5}{2}} \te[h]_k}_{L^2}^2\\
    &\qquad\qquad+\norm{\lam^{\frac{1}{2}}\te[\comp]_k}_{L^2}^2
    -2K_2(\lam^{\frac{5}{2}} \te[h]_k, \lam^{\frac{1}{2}}\te[\comp]_k)\Big]\\
    &+[\rhobar^{-1}(2\mu+\lambda)-\rhobar K_2]
      \norm{\lam^{\frac{3}{2}} \te[\comp]_k}_{L^2}^2
    +K_2\norm{\lam^{\frac{5}{2}} \te[h]_k}_{L^2}^2
   +K_2\norm{\lam^{\frac{3}{2}} \te[h]_k}_{L^2}^2\\
   =&\frac{1}{\rhobar}\Big[(\te[F]_k,\lam\te[h]_k)
   -KV'(t)\norm{\lam^{\frac{1}{2}}\te[h]_k}_{L^2}^2
    -((\vv\cdot\nabla\lam \te[h])_k,\te[h]_k)\Big]\\
   &+\frac{1}{\rhobar}\Big[(\lam \te[F]_k,\lam^2 \te[h]_k)
      -KV'(t)\norm{\lam^{\frac{3}{2}}\te[h]_k}_{L^2}^2
      -((\vv\cdot\nabla\lam \te[h])_k,\lam^2\te[h]_k)\Big]\\
   &+\rhobar^{-2}(2\mu+\lambda)K_2
    \Big[(\lam^2 \te[F]_k,\lam^3 \te[h]_k)
         -KV'(t)\norm{\lam^{\frac{5}{2}}\te[h]_k}_{L^2}^2
    -(\lam(\vv\cdot\nabla\lam
    \te[h])_k,\lam^3\te[h]_k)\Big]\\
   &+(\lam^{\frac{1}{2}}\te[G]_k,\lam^{\frac{1}{2}}\te[\comp]_k)
   -KV'(t)\norm{\lam^{\frac{1}{2}}\te[\comp]_k}_{L^2}^2
    -(\lam^{\frac{1}{2}}(\vv\cdot\nabla \te[\comp])_k,\lam^{\frac{1}{2}}\te[\comp]_k)\\
   &-K_2\Big[(\lam^2\te[F]_k,\lam \te[\comp]_k)
    +(\lam^{\frac{1}{2}} G_k,\lam^{\frac{5}{2}} h_k)
   -2KV'(t)(\lam^{\frac{5}{2}} \te[h]_k,\lam^{\frac{1}{2}}\te[\comp]_k)\\
    &\qquad\qquad\qquad-(\lam(\vv\cdot\nabla\lam \te[h])_k,
    \lam\te[\comp]_k)-(\lam(\vv\cdot\nabla\te[\comp])_k,\lam^2\te[h]_k)\Big].
\end{align*}
Noticing that
\begin{align*}
    &\abs{(\lam^{\frac{5}{2}} \te[h]_k, \lam^{\frac{1}{2}}\te[\comp]_k)}\ls
    \frac{M_3}{2}\norm{\lam^{\frac{5}{2}} \te[h]_k}_{L^2}^2
    +\frac{1}{2M_3}\norm{\lam^{\frac{1}{2}}\te[\comp]_k}_{L^2}^2,
\end{align*}
we have to choose $K_2$, $M_3$ such that
\begin{align*}
    0<K_2<M_3<\rhobar^{-2}(2\mu+\lambda).
\end{align*}
For example, we can take
\begin{align*}
    M_3=\frac{2\mu+\lambda}{2\rhobar^{2}}, \quad
    K_2=\frac{2\mu+\lambda}{4\rhobar^{2}}.
\end{align*}
Denote for $k>0$
\begin{align*}
    \alpha_k^2:=& \frac{1}{\rhobar}\norm{\lam^{\frac{1}{2}}\te[h]_k}_{L^2}^2
    +\frac{1}{\rhobar}\norm{\lam^{\frac{3}{2}} \te[h]_k}_{L^2}^2
    +\rhobar^{-2}(2\mu+\lambda)K_2\norm{\lam^{\frac{5}{2}} \te[h]_k}_{L^2}^2\\
    &+\norm{\lam^{\frac{1}{2}}\te[\comp]_k}_{L^2}^2
    -2K_2(\lam^{\frac{5}{2}} \te[h]_k, \lam^{\frac{1}{2}}\te[\comp]_k).
\end{align*}
Then, there are constants $c_1$ and $c_2$ such that
\begin{align*}
    c_1\alpha_k^2\ls\norm{\lam^{\frac{1}{2}}\te[h]_k}_{L^2}^2
    +\norm{\lam^{\frac{3}{2}} \te[h]_k}_{L^2}^2
    +\norm{\lam^{\frac{5}{2}} \te[h]_k}_{L^2}^2
    +\norm{\lam^{\frac{1}{2}}\te[\comp]_k}_{L^2}^2\ls
    c_2\alpha_k^2.
\end{align*}
Thus, by Bernstein's inequality $\norm{\lam^2 \te[h]_k}_{L^2}\ls
\frac{8}{3}2^k\norm{\lam \te[h]_k}_{L^2}$, there exists a constant
$\bar{c}$ such that
\begin{align}\label{lin4}
   \begin{aligned}
     &\frac{1}{2}\frac{d}{dt}\alpha_k^2+(\bar{c}+KV')\alpha_k^2\\
     \ls& \frac{1}{\rhobar}\Big[(\te[F]_k,\lam\te[h]_k)
    -((\vv\cdot\nabla\lam \te[h])_k,\te[h]_k)\Big]\\
   &+\frac{1}{\rhobar}\Big[(\lam \te[F]_k,\lam^2 \te[h]_k)
    -((\vv\cdot\nabla\lam
    \te[h])_k,\lam^2\te[h]_k)\Big]\\
   &+\rhobar^{-2}(2\mu+\lambda)K_2
     \Big[(\lam^{\frac{5}{2}} \te[F]_k,\lam^{\frac{5}{2}} \te[h]_k)
          -(\lam^{\frac{3}{2}}(\vv\cdot\nabla\lam
           \te[h])_k,\lam^{\frac{5}{2}}\te[h]_k)\Big]\\
   &+(\lam^{\frac{1}{2}}\te[G]_k,\lam^{\frac{1}{2}}\te[\comp]_k)
    -(\lam^{\frac{1}{2}}(\vv\cdot\nabla \te[\comp])_k,\lam^{\frac{1}{2}}\te[\comp]_k)\\
   &-K_2\Big[(\lam^{\frac{5}{2}} \te[F]_k,\lam^{\frac{1}{2}} \te[\comp]_k)
             +(\lam^{\frac{1}{2}} G_k,\lam^{\frac{5}{2}} h_k)
             -(\lam(\vv\cdot\nabla\lam \te[h])_k,\lam\te[\comp]_k)\\
   &\qquad\qquad
             -(\lam(\vv\cdot\nabla\te[\comp])_k,\lam^2\te[h]_k)\Big].
   \end{aligned}
\end{align}

Now, we combine  \eqref{lin6} and \eqref{lin4}. At this stage, we use
\lemref{lem.innner} to estimate the terms involving a convection in
\eqref{lin6} and \eqref{lin4}, and eventually get the existence of a
sequence $(\gamma_k)_{k\in\Z}$ such that $\sum_{k\in\Z}\gamma_k\ls
1$ and
\begin{align}\label{lin7}
   \begin{aligned}
   & \frac{1}{2}\frac{d}{dt}\alpha_k^2+(c\min(2^{2k},1)+KV')\alpha_k^2\\
   \ls &C\gamma_k\alpha_k2^{-k(s-1)}
   \Big[\norm{(\te[F],\te[G])}_{\hbe{s+\frac{3}{2}}\times\hbe{s-\frac{1}{2}}}
   +V'\norm{(\te[h],\te[\comp])}_{\hbe{s+\frac{3}{2}}\times\hbe{s-\frac{1}{2}}}\Big],
   \end{aligned}
\end{align}
where $c=\min(\bar{c},\hat{c})$.

We are going to show that inequality \eqref{lin7} provides us with a
decay for $h$ and $\comp$. We actually have a parabolic decay for $\comp$.

\subsection{The damping effect for $h$}
Let $\delta>0$ be a
small parameter (which will tend to $0$) and denote
$\beta_k^2=\alpha_k^2+\delta^2$. From \eqref{lin7} and dividing by
$\beta_k$, we get
\begin{align}\label{lin8}
   \begin{aligned}
   & \frac{d}{dt}\beta_k+(c\min(2^{2k},1)+KV')\beta_k\\
   \ls &C\gamma_k2^{-k(s-1)}
   \Big[\norm{(\te[F],\te[G])}_{\hbe{s+\frac{3}{2}}
    \times\hbe{s-\frac{1}{2}}}\\
   &\qquad\qquad\qquad\quad
    +V'\norm{(\te[h],\te[\comp])}_{\hbe{s+\frac{3}{2}}
     \times\hbe{s-\frac{1}{2}}}\Big]\\
   &+\delta(c\min(2^{2k},1)+KV').
   \end{aligned}
\end{align}
Integrating over $[0,t]$ and making $\delta$ tend to $0$, we have
\begin{align}\label{lin9}
   \begin{aligned}
   &\alpha_k(t)+c\min(2^{2k},1)\int_0^t\alpha_k(\tau)d\tau\\
   \ls &\alpha_k(0)+C2^{-k(s-1)}\int_0^t\gamma_k(\tau)
   \norm{(\te[F],\te[G])}_{\hbe{s+\frac{3}{2}}\times\hbe{s-\frac{1}{2}}}d\tau\\
   &+\int_0^t V'(\tau)\Big[C2^{-k(s-1)}\gamma_k(\tau)
   \norm{(\te[h],\te[\comp])}_{\hbe{s+\frac{3}{2}}
    \times\hbe{s-\frac{1}{2}}}-K\alpha_k(\tau)
   \Big]d\tau.
   \end{aligned}
\end{align}
By the definition of $\alpha_k^2$, we have for any $k\in\Z$
\begin{align}\label{lin10}
    2^{k(s-1)}\alpha_k^2\approx 2^{k(s-1)}
      \max(1,2^{\frac{5}{2}k})\norm{\te[h]_k}_{L^2}^2
    +2^{k(s-1)}\max(1,2^{\frac{k}{2}})\norm{\te[\comp]_k}_{L^2}^2.
\end{align}
Thus, we have in taking $K$ large enough such that
\begin{align*}
   \sum_{k\in\Z}\left[C2^{-k(s-1)}\gamma_k(\tau)
   \norm{(\te[h],\te[\comp])}_{\hbe{s+\frac{3}{2}}\times\hbe{s-\frac{1}{2}}}
   -K\alpha_k(\tau)\right]\ls
   0.
\end{align*}
Multiplying both sides of \eqref{lin9} by $2^{k(s-1)}$. According to
the last inequality, and due to \eqref{lin9} and \eqref{lin10},
we conclude after summation on $k$ in $\Z$, that
\begin{align}\label{lin11}
\begin{aligned}
    &\norm{\te[h](t)}_{\hbe{s+\frac{3}{2}}}
    +\norm{\te[\comp](t)}_{\hbe{s-\frac{1}{2}}}+c\int
    \norm{\te[h](\tau)}_{\hbe[s+1]{s+\frac{3}{2}}}\\
    &\qquad\qquad+\sum_{k\in\Z}\int_0^t
    c2^{k(s-\frac{1}{2})}\min(2^{\frac{3k}{2}},1)\norm{\te[\comp]_k(\tau)}_{L^2}d\tau\\
    \ls &\norm{(\te[h](0),\te[\comp](0))}_{\hbe{s+\frac{3}{2}}
    \times\hbe[s-1]{s-\frac{1}{2}}}\\
    &\qquad\qquad+\int_0^t
    \norm{(\te[F],\te[G])(\tau)}_{\hbe{s+\frac{3}{2}}
    \times\times\hbe[s-1]{s-\frac{1}{2}}}d\tau.
    \end{aligned}
\end{align}

\subsection{The smoothing effect for $\comp$}

Once stated the
damping effect for $h$, it is easy to get the smoothing effect on
$\comp$. Since \eqref{lin11} implies the desired estimate for low
frequencies, it suffices to prove it for high frequencies only. We
therefore suppose in this part that $k>0$.

Define $\theta_k=\norm{\te[\comp]_k}_{L^2}$. By the previous
inequalities and using \lemref{lem.innner}, the second equality of
\eqref{lin2} yields, for a constant $c>0$, that
\begin{align*}
    \frac{1}{2}\frac{d}{dt}\theta_k^2+c2^{2k}\theta_k^2
    \lesssim & \theta_k(\norm{\lam^2\te[h]_k}_{L^2}+\norm{\te[G]_k}_{L^2})\\
    &+\theta_kV'(t)(C\gamma_k2^{-k(s-1)}
    \min(1,2^{-\frac{k}{2}})
    \norm{\te[\comp]}_{\hbe[s-1]{s-\frac{1}{2}}}-K\theta_k).
\end{align*}
Using $\beta_k^2=\theta_k^2+\delta^2$, integrating over $[0,t]$ and
then having $\delta$ tend to $0$, we infer
\begin{align*}
    \theta_k(t)+c2^{2k}\theta_k\ls& \theta_k(0)
    +C\int_0^t\norm{\te[G]_k}_{L^2}d\tau
    +C\int_0^t2^{2k}\norm{\te[h]_k(\tau)}_{L^2}d\tau\\
    &+C\int_0^t
    V'(\tau)\gamma_k(\tau)2^{-k(s-1)}\min(1,2^{-\frac{k}{2}})
    \norm{\te[\comp](\tau)}_{\hbe[s-1]{s-\frac{1}{2}}}d\tau.
\end{align*}
Therefore, we get
\begin{align*}
    &\sum_{k>0} 2^{k(s-\frac{1}{2})}\norm{\te[\comp]_k(t)}_{L^2}
    +c\int_0^t \sum_{k>0}
    2^{k(s+\frac{3}{2})}\norm{\te[\comp]_k(\tau)}_{L^2}d\tau\\
    \ls& \norm{\te[\comp](0)}_{\hbe[s-1]{s-\frac{1}{2}}}
    +C\int_0^t\norm{\te[G](\tau)}_{\hbe[s-1]{s-\frac{1}{2}}}d\tau
    +C\int_0^t\sum_{k>0}2^{k(s+\frac{3}{2})}\norm{\te[h]_k(\tau)}_{L^2}d\tau\\
    &\qquad +CV(t)\sup_{[0,t]}\norm{\te[\comp](t)}_{\hbe[s-1]{s-\frac{1}{2}}}.
\end{align*}
Using \eqref{lin11}, we eventually conclude that
\begin{align*}
    &c\int_0^t \sum_{l>0}
    2^{k(s+\frac{3}{2})}\norm{\te[\comp]_k(\tau)}_{L^2}d\tau\\
    \ls&
    (C+CV(t))\Big(\norm{\te[h](0)}_{\hbe{s+\frac{3}{2}}}
    +\norm{\te[\comp](0)}_{\hbe[s-1]{s-\frac{1}{2}}})\\
    &\qquad\qquad\qquad
    +\int_0^t(\norm{\te[F](\tau)}_{\hbe{s+\frac{3}{2}}}
    +\norm{\te[G](\tau)}_{\hbe[s-1]{s-\frac{1}{2}}})d\tau
    \Big).
\end{align*}
Combining the last inequality with \eqref{lin11}, we complete the
proof of Proposition \ref{prop.lin} as long as we change the
functions $(\te[h],\te[\comp],\te[F],\te[G])$ back into the original
ones $(h,\comp,F,G)$.

\section{A global existence and uniqueness result}
\label{sec.proof}
This section is devoted to the proof of \thmref{thm.main}. The
principle of the proof is a very classical one. We shall use the
classical Friedrichs' regularization method, which was used in
\cite{CheM04,CheZ07,Paicu} for examples, to construct the
approximate solutions $(h^n,\vu^n)$ of \eqref{nsp4}.

\subsection{Building of the sequence $(h^n,\vu^n)_{n\in\NN}$}
Let us define the sequence of operators $(\fr )_{n\in\NN}$ by
\begin{align*}
    \fr  f:=\F^{-1}\mathbf{1}_{B(\frac{1}{n},n)}(\xi)\F f.
\end{align*}
We consider the approximate system:
\begin{align}\label{bui.1}
\left\{\begin{aligned}
    &h_t^{n}+\fr \lam^{-1}(\fr \vu^n\cdot\nabla \lam \fr  h^{n})
      + \rhobar \fr  \comp^{n}=F^n,\\
    &\comp_t^{n}+\fr (\fr \vu^n\cdot\nabla \fr \comp^{n})
      -\rhobar^{-1}(2\mu+\lambda)\Delta \fr  \comp^{n}\\
      &\qquad\qquad\qquad\qquad\qquad\qquad-\lam^2 \fr  h^{n}
    -\fr h^{n}=G^n,\\
    &\inc_t^{n}-\rhobar^{-1}\mu\Delta \fr \inc^{n}=H^n,\\
    &\vu^{n}=-\lam^{-1}\nabla \comp^{n}-\lam^{-1}\dive\inc^{n},\\
    &(h^{n},\comp^{n},\inc^{n})(0)=(h_n,\lam^{-1}\dive\vu_n,
    \lam^{-1}\curl\vu_n),
    \end{aligned}\right.
\end{align}
where
\begin{align*}
    & h_n=\fr (\rho_0-\rhobar), \quad
    \vu_n=\fr \vu_0,\\
    &F^n=-\fr \lam^{-1}(\lam \fr  h^n\dive \fr \vu^n),\\
    &G^n=\fr (\fr \vu^n\cdot\nabla \fr  \comp^n)-\fr \lam^{-1}\dive J^n,\\
    &H^n=-\fr \lam^{-1}\curl J^n,\\
    &J^n=\fr (\fr \vu^n\cdot\nabla \fr \vu^n)
      +\frac{\lam \fr h^n}{\rhobar \zeta(\lam \fr h^n+\rhobar)}
    (\mu\Delta \fr \vu^n+(\mu+\lambda)\nabla\dive \fr \vu^n),
\end{align*}
where $\zeta$ is a smooth function satisfying
\begin{align*}
    \zeta(s)=\left\{\begin{array}{ll}
        \rhobar/4,&\abs{s}\ls\rhobar/4,\\
        s, &\rhobar/2\ls\abs{s}\ls 3\rhobar/2,\\
        7\rhobar/4, &\abs{s}\gs 7\rhobar/4.\\
        \text{smooth}, &\text{otherwise}.
    \end{array}\right.
\end{align*}

We want to show that \eqref{bui.1} is only an ordinary differential
equation in $L^2\times L^2\times L^2$. We can observe easily that
all the source term in \eqref{bui.1} turn out to be continuous in
$L^2\times L^2\times L^2$. For example, we consider the term $\fr
\lam^{-1}\dive \frac{\lam \fr  h^n\Delta \fr \vu^n}{\zeta(\lam \fr
h^n+\rhobar)}$. By Plancherel's theorem, Hausdorff-Young's
inequality and H\"older's inequality, we have
\begin{align*}
   &\norm{\fr \lam^{-1}
    \dive \frac{\lam \fr  h^n\Delta \fr \vu^n}
               {\zeta(\lam \fr   h^n+\rhobar)}}_{L^2}
   =\norm{\mathbf{1}_{B(\frac{1}{n},n)}
    \abs{\xi}^{-1}\xi\cdot\F
     \frac{\lam \fr  h^n\Delta \fr \vu^n}
          {\zeta(\lam \fr   h^n+\rhobar)}}_{L^2}
\\
   \ls &\norm{\frac{\lam \fr  h^n\Delta \fr \vu^n}
                   {\zeta(\lam \fr   h^n+\rhobar)}}_{L^2}
   \ls \norm{\lam \fr  h^n\Delta \fr \vu^n}_{L^2}
       \norm{\frac{1}{\zeta(\lam \fr   h^n+\rhobar)}}_{L^\infty}
\\
   \ls&\frac{4}{\rhobar}
       \norm{\lam \fr  h^n}_{L^\infty}
       \norm{\Delta \fr \vu^n}_{L^2}
   \ls \frac{4n^2}{\rhobar}
     \norm{\abs{\xi}\mathbf{1}_{B(\frac{1}{n},n)}\F h^n}_{L^1}
     \norm{\vu^n}_{L^2}\\
   \lesssim
   &\frac{4n^{\frac{N}{2}+3}}{\rhobar}
     \norm{h^n}_{L^2}\norm{\vu^n}_{L^2}.
\end{align*}
Thus, the usual Cauchy-Lipschitz theorem implies the existence of a
strictly positive maximal time $T_n$ such that a unique solution
exists which is continuous in time with value in $L^2\times
L^2\times L^2$. However, as $\fr ^2=\fr $, we claim that $\fr
(h^n,\comp^n,\inc^n)$ is also a solution, so uniqueness implies that
$\fr (h^n,\comp^n,\inc^n)=(h^n,\comp^n,\inc^n)$. So $(h^n,\comp^n,\inc^n)$ is
also a solution of the following system:
\begin{align}\label{bui.2}
\left\{\begin{aligned}
    &h_t^{n}+\fr \lam^{-1}(\vu^n\cdot\nabla \lam  h^{n})+ \rhobar  \comp^{n}=F_1^n,\\
    &\comp_t^{n}+\fr (\vu^n\cdot\nabla \comp^{n})
            -\rhobar^{-1}(2\mu+\lambda)\Delta \comp^{n}-\lam^2  h^{n}
            -h^{n}=G_1^n,\\
    &\inc_t^{n}-\rhobar^{-1}\mu\Delta \inc^{n}=H_1^n,\\
    &\vu^{n}=-\lam^{-1}\nabla \comp^{n}-\lam^{-1}\dive\inc^{n},\\
    &(h^{n},\comp^{n},\inc^{n})(0)=(h_n,\lam^{-1}\dive\vu_n,
    \lam^{-1}\curl\vu_n),
    \end{aligned}\right.
\end{align}
where
\begin{align*}
    &F_1^n=-\fr \lam^{-1}(\lam  h^n\dive \vu^n),\\
    &G_1^n=\fr (\vu^n\cdot\nabla  \comp^n)-\fr \lam^{-1}\dive J^n,\\
    &H_1^n=-\fr \lam^{-1}\curl J^n,\\
    &J_1^n=\fr (\vu^n\cdot\nabla \vu^n)
      +\frac{\lam h^n}{\rhobar \zeta(\lam h^n+\rhobar)}
    (\mu\Delta \vu^n+(\mu+\lambda)\nabla\dive \vu^n).
\end{align*}
The system \eqref{bui.2} appears to be an ordinary differential
equation in the space
\begin{align*}
    L_n^2:=\set{a\in L^2(\Real^N): \supp\F a\subset B(\frac{1}{n},n)}.
\end{align*}
Due to the Cauchy-Lipschitz theorem again, a unique maximal
solution exists on an interval $[0,T_n^*)$ which is continuous in
time with value in $L_n^2\times L_n^2\times L_n^2$.

\subsection{Uniform bounds}
In this part, we prove uniform estimates independent of $T<T_n^*$ in
$\en$ for $(h^n,\vu^n)$. We shall show that $T_n^*=+\infty$ by the
Cauchy-Lipschitz theorem. Define
\begin{align*}
    E(0):=&\norm{\lam^{-1}(\rho_0-\rhobar)}_{\hbe[\hn-\frac{3}{2}]{\hn+1}}
    +\norm{\vu_0}_{\hbe[\hn-\frac{3}{2}]{\hn-1}},\\
    E(h,\vu,t):=&\norm{(h,\vu)}_{E_t^\hn},\\
    \te[T]_n:=&\sup\set{t\in[0,T_n^*): E(h^n,\vu^n,t)\ls A\te[C]E(0)},
\end{align*}
where $\te[C]$ corresponds to the constant in \propref{prop.lin} and
$A>\max(2,\te[C]^{-1})$ is a constant. Thus, by the continuity we
have $\te[T]_n>0$.

We are going to prove that $\te[T]_n=T_n^*$ for all $n\in\NN$ and we
will conclude that $T_n^*=+\infty$ for any $n\in\NN$.

According to \propref{prop.lin} and \lemref{lem.heat}, and to the
definition of $(h_n,\vu_n)$, the following inequality holds
\begin{align*}
    \norm{(h^{n},\vu^{n})}_\ent
\ls &\te[C]e^{\te[C]\norm{\vu^n}_{L_T^1(\be[\hn+1])}}
    \Big(\norm{\lam^{-1}(\rho_0-\rhobar)}_{\hbe[\hn-\frac{3}{2}]{\hn+1}}
\\
    &+\norm{\vu_0}_{\hbe[\hn-\frac{3}{2}]{\hn-1}}
     +\norm{F_1^n}_{L_T^1(\hbe[\hn-\frac{3}{2}]{\hn+1})}\\
    &+\norm{\vu^n\cdot\nabla \comp^n}_{L_T^1(\hbe[\hn-\frac{3}{2}]{\hn-1})}
    +\norm{J_1^n}_{L_T^1(\hbe[\hn-\frac{3}{2}]{\hn-1})}\Big).
\end{align*}

Therefore, it is only a matter of proving appropriate estimates for
$F_1^n$, $J_1^n$ and the convection term. The estimate of $F_1^n$ is
straightforward. From \lemref{lem.fgbesov}, we have
\begin{align}\label{uni.2}
    \begin{aligned}
    \norm{F_1^n}_{L_T^1(\hbe[\hn-\frac{3}{2}]{\hn+1})}
    =&\norm{\lam h^n\dive\vu^n}_{L_T^1(\hbe[\hn-\frac{5}{2}]{\hn})}\\
    \ls &C\norm{\lam h^n}_{L_T^\infty(\hbe[\hn-2]{\hn})}
    \norm{\dive\vu^n}_{L_T^1(\hbe[\hn-\frac{1}{2}]{\hn})}\\
    \ls &C\norm{h^n}_{L_T^\infty(\hbe[\hn-\frac{3}{2}]{\hn+1})}
    \norm{\vu^n}_{L_T^1(\hbe[\hn+\frac{1}{2}]{\hn+1})}\\
    \ls &CE^2(h^n,\vu^n,T).
    \end{aligned}
\end{align}
With the help of \lemref{lem.fgbesov} and interpolation arguments, we have
\begin{align}\label{uni.3}
    \begin{aligned}
    &\norm{\vu^n\cdot\nabla \comp^n}_{L_T^1(\hbe[\hn-\frac{3}{2}]{\hn-1})}\\
    \ls &C\norm{\vu^n}_{L_T^\infty(\hbe[\hn-1]{\hn-1})}
    \norm{\nabla \comp^n}_{L_T^1(\hbe[\hn-\frac{1}{2}]{\hn})}\\
    \ls &C\norm{\vu^n}_{L_T^\infty(\hbe[\hn-\frac{3}{2}]{\hn-1})}
    \norm{\vu^n}_{L_T^1(\hbe[\hn+\frac{1}{2}]{\hn+1})}\\
    \ls &CE^2(h^n,\vu^n,T).
    \end{aligned}
\end{align}
In the same way, we can get
\begin{align}\label{uni.4}
\norm{\vu^n\cdot\nabla \vu^n}_{L_T^1(\hbe[\hn-\frac{3}{2}]{\hn-1})}
 \ls CE^2(h^n,\vu^n,T).
\end{align}
To estimate other terms of $J_1^n$, we make the following assumption on $E(0)$:
\begin{align}\label{uni.5}
    2C_1A\te[C]E(0)\ls \rhobar,
\end{align}
where $C_1$ is the continuity modulus of the embedding relation
$\be[\hn](\Real^N)\hookrightarrow L^\infty(\Real^N)$. If
$T<\te[T]_n$, it implies
\begin{align}\label{uni.6}
\begin{aligned}
    \norm{\lam h^n}_{L^\infty([0,T]\times\Real^N)}
\ls &C_1\norm{h^n}_{L_T^\infty(\be[\hn+1])}
    \ls C_1\norm{h^n}_{L_T^\infty(\hbe[\hn-\frac{3}{2}]{\hn+1})}\\
    \ls &C_1A\te[C]E(0) \ls \frac{1}{2}\rhobar,
    \end{aligned}
\end{align}
which yields
\begin{align*}
    \lam h^n+\rhobar\in[\frac{1}{2}\rhobar,\frac{3}{2}\rhobar]
    \text{ and } \zeta(\lam h^n+\rhobar)=\lam h^n+\rhobar.
\end{align*}

From \lemref{lem.fgbesov} and \lemref{lem.comp}, we obtain
\begin{align}\label{uni.7}
    \begin{aligned}
    &\norm{\frac{\lam h^n}{\rhobar (\lam h^n+\rhobar)}
    (\mu\Delta \vu^n+(\mu+\lambda)\nabla\dive \vu^n)
     }_{L_T^1(\hbe[\hn-\frac{3}{2}]{\hn-1})}\\
    \ls &C\norm{\frac{\lam h^n}{\lam h^n+\rhobar}}_{L_T^\infty(\be[\hn])}
    \norm{\vu^n}_{L_T^1(\hbe[\hn+\frac{1}{2}]{\hn+1})}\\
    \ls &C\norm{\lam h^n}_{L_T^\infty(\be[\hn])}
         \norm{\vu^n}_{L_T^1(\hbe[\hn+\frac{1}{2}]{\hn+1})}\\
    \ls &C\norm{ h^n}_{L_T^\infty(\hbe[\hn-\frac{3}{2}]{\hn+1})}
    \norm{\vu^n}_{L_T^1(\hbe[\hn+\frac{1}{2}]{\hn+1})}\\
    \ls &CE^2(h^n,\vu^n,T).
    \end{aligned}
\end{align}
Thus, we get
\begin{align*}
    \norm{(h^{n},\vu^{n})}_\ent
\ls &\te[C]e^{A\te[C]^2E(0)}[1+CA^2\te[C]^2E(0)]E(0).
\end{align*}
So we can choose $E(0)$ so small that
\begin{align}\label{uni.8}
1+CA^2\te[C]^2E(0)
    \ls \frac{A^2}{A+2},\quad e^{A\te[C]E(0)}
    \ls \frac{A+1}{A},\quad 2C_1A\te[C]E(0)\ls \rhobar,
\end{align}
which yields $\norm{(h^{n},\vu^{n})}_{E_T^1}\ls
\frac{A+1}{A+2}A\te[C]E(0)$ for any $T<\te[T]_n$. It follows that
$\te[T]_n=T_n^*$. In fact, if $\te[T]_n<T_n^*$, we have seen that
$E(h^n,\vu^n,\te[T]_n)\ls \frac{A+1}{A+2}A\te[C]E(0)$. So by
continuity, for a sufficiently small constant $\sigma>0$ we can
obtain $E(h^n,\vu^n,\te[T]_n+\sigma)\ls A\te[C]E(0)$. This yields a
contradiction with the definition of $\te[T]_n$.

Now, if $\te[T]_n=T_n^*<\infty$, we have obtained
$F(h^n,\vu^n,T_n^*)\ls A\te[C]E(0)$. As
$\norm{h^n}_{L_{T_n^*}(\hbe[0]{1+\eps})}<\infty$ and
$\norm{\vu^n}_{L_{T_n^*}(\hbe[0]{\eps})}<\infty$, it implies that
$\norm{h^n}_{L_{T_n^*}(L_n^2)}<\infty$ and
$\norm{\vu^n}_{L_{T_n^*}(L_n^2)}<\infty$. Thus, we may continue the
solution beyond $T_n^*$ by the Cauchy-Lipschitz theorem. This
contradicts the definition od $T_n^*$. Therefore, the approximate
solution $(h^n,\vu^n)_{n\in\NN}$ is global in time.

\subsection{Existence of a solution}

In this part, we shall show that, up to an extraction, the sequence
$(h^n,\vu^n)_{n\in\NN}$ converges in
$\mathscr{D}'(\Real^+\times\Real^N)$ to a solution $(h,\vu)$ of
\eqref{nsp4} which has the desired regularity properties. The proof
lies on compactness arguments. To start with, we show that the time
first derivative of $(h^n,\vu^n)$ is uniformly bounded in
appropriate spaces. This enables us to apply Ascoli's theorem and
get the existence of a limit $(h,\vu)$ for a subsequence. Now,
the uniform bounds of the previous part provides us with additional
regularity and convergence properties so that we may pass to the
limit in the system.

It is convenient to split $(h^n,\vu^n)$ into the solution of a
linear system with initial data $(h_n,\vu_n)$, and the
discrepancy to that solution. More precisely, we denote by
$(h_L^n,\vu_L^n)$ the solution to the linear system
\begin{align}\label{ex.1}
\left\{\begin{aligned}
    &\partial_t h_L^n+\rhobar\lam^{-1}\dive\vu_L^n=0,\\
    &\partial_t\vu_L^n-\rhobar^{-1}\mu\Delta\vu_L^n
    -\rhobar^{-1}(\mu+\lambda)\nabla\dive\vu^n+\nabla\lam h_L^n
    +\nabla\lam^{-1} h_L^n=0,\\
    &(h_L^n,\vu_L^n)_{t=0}=(h_n,\vu_n),
    \end{aligned}\right.
\end{align}
and $(\bar{h}^n,\bar{\vu}^n)=(h^n-h_L^n,\vu^n-\vu_L^n)$.

Obviously, the definition of $(h_n,\vu_n)$ entails
\begin{align*}
    h_n\to \lam^{-1}(\rho_0-\rhobar)
     \text{ in } \hbe[\hn-\frac{3}{2}]{\hn+1}, \quad
    \vu_n\to\vu_0
     \text{ in } \hbe[\hn-\frac{3}{2}]{\hn-1} \text{ as } n\to\infty.
\end{align*}
The \lemref{lem.heat} and \propref{prop.lin} insure us
that
\begin{align}\label{ex.2}
    (h_L^n,\vu_L^n) \to (h_L,\vu_L) \text{ in } E^\hn,
\end{align}
where $(h_L,\vu_L)$ is the solution of the linear system
\begin{align}\label{ex.3}
\left\{\begin{aligned}
    &\partial_t h_L+\rhobar\lam^{-1}\dive\vu_L=0,\\
    &\partial_t\vu_L-\rhobar^{-1}\mu\Delta\vu_L
    -\rhobar^{-1}(\mu+\lambda)\nabla\dive\vu+\nabla\lam h_L
    +\nabla\lam^{-1} h_L=0,\\
    &(h_L,\vu_L)_{t=0}=(\lam^{-1}(\rho_0-\rhobar),\vu_0).
    \end{aligned}\right.
\end{align}

Now, we have to prove the convergence of
$(\bar{h}^n,\bar{\vu}^n)$. This is of course a trifle more
difficult and requires compactness results. Let us first state the
following lemma.

\begin{lem}\label{lem.ubt}
$((\bar{h}^n,\bar{\vu}^n))_{n\in\NN}$ is uniformly bounded in
\begin{align*}
\C^{\frac{1}{2}}(\Real^+; \be[\hn-\frac{3}{2}])
 \times (\C^{\frac{1}{4}}(\Real^+; \be[\hn-\frac{3}{2}]))^N.
\end{align*}
\end{lem}

\begin{pf}
Throughout the proof, we will note u.b. for uniformly bounded. We
first prove that $\partial_t \bar{h}^n$ is u.b. in
$(L^2+L^\infty)(\Real^+,\be[\hn-\frac{3}{2}])$, which yields the
desired result for $\bar{h}$. Let us observe that $\bar{h}^n$
verifies the following equation
\begin{align*}
    \partial_t\bar{h}^{n}=&-\fr \lam^{-1}(\lam h^n\dive\vu^n)
    -\fr \lam^{-1}(\vu^n\cdot\nabla\lam h^{n})\\
    &-\rhobar\lam^{-1}\dive\vu^{n}+\rhobar\lam^{-1}
    \dive\vu_L^{n}.
\end{align*}
According to the previous part, $(h^n)_{n\in\NN}$ is u.b. in
$L^\infty(\be[\hn-\frac{1}{2}])$ and $(\vu^n)_{n\in\NN}$ is u.b. in
$L^2(\be[\hn])$ in view of interpolation arguments. Thus, $\fr
\lam^{-1}(\lam h^n\dive\vu^n)$, $\fr \lam^{-1}(\vu^n\cdot\nabla\lam
h^{n})$, $\rhobar\lam^{-1}\dive\vu^{n}$ is u.b. in
$L^2(\be[\hn-\frac{3}{2}])$. The definition of $\vu_L^n$ obviously
provides us with uniform bounds for $\lam^{-1}\dive\vu_L^n$ in
$L^\infty(\be[\hn-\frac{3}{2}])$, so we can conclude that
$\partial_t\bar{h}^n$ is u.b. in
$(L^2+L^\infty)(\be[\hn-\frac{3}{2}])$.

Denote $\comp_L^n=\lam^{-1}\dive\vu_L^n$, $\bar{\comp}^n=
\lam^{-1}\dive\bar{\vu}^n$, $\inc_L^n= \lam^{-1}\curl\vu_L^n$ and
$\bar{\inc}^n= \lam^{-1}\curl\bar{\vu}^n$. Let us prove now that
$\partial_t\bar{\comp}^n$ is u.b. in
$(L^{\frac{4}{3}}+L^\infty)(\Real^+;\be[\hn-\frac{3}{2}])$ and that
$\partial_t\bar{\inc}^n$ is u.b. in $L^\frac{4}{3}(\Real^+;
\be[\hn-\frac{3}{2}])$ which give the required result for
$\bar{\vu}^n$ by using the relation $\vu^n=-\lam^{-1}\nabla
\comp^n-\lam^{-1}\dive\inc^n$.

Let us recall that
\begin{align*}
    \partial_t\bar{\comp}^{n}=&\rhobar^{-1}(2\mu+\lambda)\Delta(\comp^{n}-\comp_L^{n})
    +\lam^2(h^{n}-h_L^{n})\\
    &+(h^{n}-h_L^{n})
    -\fr \lam^{-1}\dive J^n,\\
    \partial_t\bar{\inc}^{n}=&\rhobar^{-1}\mu\Delta(\inc^{n}-\inc_L^{n})
    -\fr \lam^{-1}\curl J^n.
\end{align*}
Results of the previous part and an interpolation argument yield
uniform bounds for $\vu^n$ and $\comp^n$ in
$L^{\frac{4}{3}}(\be[\hn+\frac{1}{2}])\cap L^2(\be[\hn])$. Since
$h^n$ is u.b. in $L^\infty(\be[\hn+1])$ and $\comp_L^n$ is u.b. in
$L^{\frac{4}{3}}(\be[\hn+\frac{1}{2}])$, we easily verify that
$\Delta(\comp^{n}-\comp_L^{n})$ and $\fr \lam^{-1}\dive J^n$ are u.b. in
$L^{\frac{4}{3}}(\be[\hn-\frac{3}{2}])$. Because $h^n$ is u.b. in
$L^\infty(\be[\hn+\frac{1}{2}])$, $\lam^2 h^n$ is u.b. in
$L^\infty(\be[\hn-\frac{3}{2}])$. We also have $\lam^2h_L^n$ u.b. in
$L^\infty(\be[\hn-\frac{3}{2}])$. In addition, $h^n$ and $h_L^n$ are
u.b. in $L^\infty(\be[\hn-\frac{3}{2}])$. So we finally get
$\partial_t\bar{\comp}^n$ u.b. in
$(L^{\frac{4}{3}}+L^\infty)(\Real^+;\be[\hn-\frac{3}{2}])$. The case
of $\partial_t\bar{\inc}^n$ goes along the same lines. As the terms
corresponding to $(h^n-h_L^n)$ do not appear, we simply get
$\partial_t\bar{\inc}^n$ u.b. in
$L^{\frac{4}{3}}(\be[\hn-\frac{3}{2}])$.
\end{pf}

Now, we can turn to the proof of the existence of a solution and use
Ascoli theorem to get strong convergence. We need to localize the
spatial space because we have some results of compactness for the
local Sobolev spaces. Let $(\chi_p)_{p\in\NN}$ be a sequence of
$\C_0^\infty(\Real^N)$ cut-off functions supported in the ball
$B(0,p+1)$ of $\Real^N$ and equal to $1$ in a neighborhood of
$B(0,p)$.

For any $p\in\NN$, Lemma~\ref{lem.ubt} tells us that
$((\chi_p\bar{\rho}^n, \chi_p\bar{\vu}^n))_{n\in\NN}$ is uniformly
equicontinuous in $\C(\Real^+;(\be[\hn-\frac{3}{2}])^{1+N})$.

Let us observe that the application $f\mapsto \chi_p f$ is compact
from $\hbe[\hn-\frac{3}{2}]{\hn+1}$ into $\be[\hn-\frac{3}{2}]$, and
from $\hbe[\hn-\frac{3}{2}]{\hn-1}$ into $\be[\hn-\frac{3}{2}]$.
After we apply Ascoli's theorem to the family $((\chi_p\bar{h}^n,
\chi_p\bar{\vu}^n))_{n\in\NN}$ on the time interval $[0,p]$, we use
Cantor's diagonal process. This finally provides us with a
distribution $(\bar{h},\bar{\vu})$ belonging to $\C(\Real^+;
(\be[\hn-\frac{3}{2}])^{1+N})$ and a subsequence (which we still
denote by $((\bar{\rho}^n,\bar{\vu}^n)_{n\in\NN})$ such that, for
all $p\in\NN$, we have
\begin{align}\label{ex.4}
    (\chi_p\bar{h}^n,\chi_p\bar{\vu}^n)\to
    (\chi_p\bar{h},\chi_p\bar{\vu}) \text{ as } n\to+\infty, \text{
    in } \C([0,p]; (\be[\hn-\frac{3}{2}])^{1+N}).
\end{align}
This obviously infers that $(\bar{h}^n,\bar{\vu}^n)$ tends to
$(\bar{h},\bar{\vu})$ in $\mathscr{D}'(\Real^+\times \Real^N)$.

Coming back to the uniform estimates of the previous part, we
moreover get that $(\bar{h},\bar{\vu})$ belongs to
\begin{align*}
    &L^\infty\left(\Real^+;\hbe[\hn-\frac{3}{2}]{\hn+1}
    \times(\hbe[\hn-\frac{3}{2}]{\hn-1})^N\right)\cap L^1\left(\Real^+;
    (\hbe[\hn+\frac{1}{2}]{\hn+1})^{1+N}\right)
\end{align*}
and to $\C^{1/2}(\Real^+;\be[\hn-\frac{3}{2}])\times
(C^{1/4}(\Real^+;\be[\hn-\frac{3}{2}]))^N$.

Let us now prove that
$(h,\vu):=(h_L,\vu_L)+(\bar{h},\bar{\vu})$ solves
\eqref{nsp4}. We first observe that, according to \eqref{bui.1},
\begin{align}\label{ex.5}
\left\{\begin{aligned}
  &h_t^{n}+\fr \lam^{-1}(\vu^n\cdot\nabla\lam h^{n})
          +\rhobar \comp^{n}=-\fr \lam^{-1}(\lam h^n\dive\vu^n),
\\
 &\vu_t^{n}+\fr (\vu^n\cdot\nabla\vu^n)-\rhobar^{-1}\mu\Delta\vu^{n}
   -\rhobar^{-1}(\mu+\lambda) \nabla\dive\vu^{n}
   +\lam\nabla h^{n}\\
 &\qquad+\lam^{-1}\nabla h^{n}=-\fr \frac{ \lam h^n}{\rhobar(\lam h^n+\rhobar)}
    (\mu\Delta \vu^n+(\mu+\lambda)\nabla\dive \vu^n).
\end{aligned}\right.
\end{align}

The only problem is to pass to the limit in
$\mathscr{D}'(\Real^+\times\Real^N)$ in nonlinear terms. This
can be done by using the convergence results stemming from the
uniform estimates and the convergence results \eqref{ex.2} and
\eqref{ex.3}.

As it is just a matter of doing tedious verifications, we show, as
an example, the case of the term
$\fr \frac{\lam h^n\Delta\vu^n}{\rhobar(\lam h^n+\rhobar)}$. Denote
$L(z)=z/(z+\rhobar)$. Let $\theta\in\C_0^\infty(\Real^+\times\Real^N)$ and
$p\in\NN$ be such that $\supp\theta\subset [0,p]\times B(0,p)$. We consider the decomposition
\begin{align*}
    &\fr\frac{\theta\lam h^n\Delta\vu^n}{\rhobar(\lam h^n+\rhobar)}
    -\frac{\theta\lam h\Delta\vu}{\rhobar(\lam h+\rhobar)}\\
    =&\rhobar^{-2}\fr[\theta(1-L(\lam h^n))\chi_p\lam h^n\chi_p
    \Delta(\vu_L^n-\vu_L)\\
    &+\theta(1-L(\lam h^n))
    \chi_p\lam h^n\chi_p\Delta(\chi_p(\bar{\vu}^n-\bar{\vu}))\\
    & +\theta(1-L(\lam h^n))
    (\chi_p\lam(h^n-h))\Delta\vu
    -\theta\lam h\chi_p\Delta\vu(L(\chi_p\lam h^n)-L(\chi_p\lam  h))]\\
    &+(\fr-I)\frac{\theta\lam h\Delta\vu}{\rhobar(\lam h+\rhobar)}.
\end{align*}
The last term tends to zero as $n\to +\infty$ due to the property of
$\fr$. As $\theta L(\lam h^n)$ and $\lam h^n$ are u.b. in
$L^\infty(\be[\hn])$ and $\vu_L^n$ tends to $\vu_L$ in
$L^{\frac{4}{3}}(\be[\hn-\frac{3}{2}])$, the first term tends to $0$
in $L^{\frac{4}{3}}(\be[\hn-\frac{3}{2}])$. According to
\eqref{ex.3}, $\chi_p(\bar{\vu}^n-\bar{\vu})$ tends to zero in
$L^{\frac{4}{3}}(\be[\hn-\frac{3}{2}])$ so that the second term
tends to $0$ in $L^{\frac{4}{3}}(\be[\hn-\frac{3}{2}])$. Clearly,
$\chi_p \lam h^n\to\chi_p\lam h$ in $L^\infty(\be[\hn])$ and
$L(\chi_p\lam h^n)\to L(\chi_p\lam h)$ in
$L^\infty(L^\infty\cap\be[\hn])$, so that the third and the last
terms also tend to $0$ in $L^{\frac{4}{3}}(\be[\hn-\frac{3}{2}])$.
The other nonlinear terms can be treated in the same way.

We still have to prove that $h$ is continuous in
$\hbe[\hn-\frac{3}{2}]{\hn+1}$ and that $\vu$ belongs to
$\C(\Real^+;\hbe[\hn-\frac{3}{2}]{\hn-\frac{1}{2}})$. The continuity
of $\vu$ is straightforward. Indeed, $\vu$ satisfies
\begin{align*}
 \vu_t=&-\vu\cdot\nabla\vu+\rhobar^{-1}\mu\Delta\vu
          +\rhobar^{-1}(\mu+\lambda) \nabla\dive\vu\\
    &-\lam\nabla h-\lam^{-1}\nabla h
       -\frac{\lam h}{\rhobar (\lam h+\rhobar)}
    (\mu\Delta \vu+(\mu+\lambda)\nabla\dive \vu)
\end{align*}
and the r.h.s. belongs to
$(L^1+L^\infty)(\Real^+;\hbe[\hn-\frac{3}{2}]{\hn-1})$. We have
already got that $h\in\C(\Real^+;\be[\hn-\frac{3}{2}])$. Indeed,
$h_t\in L^\infty(\Real^+;\be[\hn-\frac{3}{2}])$ from the equation
\begin{align*} h_t=-\lam^{-1}\dive(\lam
h\vu)-\rhobar\lam^{-1}\dive\vu.
\end{align*}
Thus, there remains to prove the continuity of $h$ in $\be[\hn+1]$.

Let us apply the operator $dk$ to the first equation of \eqref{nsp4} to get
\begin{align}\label{ex.6}
    \partial_t\dk \lam h
 =-\dk(\vu\cdot\nabla\lam h)-\rhobar \dk\dive \vu
  -\dk(\lam h\dive\vu).
\end{align}
Obviously, for fixed $k$ the r.h.s. belongs to $L_{loc}^1(\Real^+;
L^2)$ so that each $\dk \lam h$ is continuous in time with values in
$L^2$.

Now, we apply an energy method to \eqref{ex.4} to obtain, with the
help of \lemref{lem.innner}, that
\begin{align*}
    \frac{1}{2}\frac{d}{dt}\norm{\dk\lam h}_{L^2}^2
 \ls C \norm{\dk \lam h}_{L^2}
    \Big(&\gamma_k 2^{-k\hn}\norm{\lam h}_{\be[\hn]}
          \norm{\vu}_{\be[\hn+1]}+\norm{\dk\dive\vu}_{L^2}\\
         &+\norm{\dk(\lam h\dive\vu)}_{L^2}\Big),
\end{align*}
where $\sum_{k\in\Z}\gamma_k\ls 1$. Integrating in time and
multiplying $2^{k\hn}$, we get
\begin{align*}
    2^{k(\hn+1)}\norm{\dk h(t)}_{L^2}
 \ls &2^{k(\hn+1)}\norm{\dk\lam^{-1}(\rho_0-\rhobar)}_{L^2}\\
    &+C\int_0^t\Big(\gamma_k\norm{h(\tau)}_{\be[\hn+1]}
    \norm{\vu(\tau)}_{\be[\hn+1]}\\
    &+2^{k(\hn+1)}\norm{\dk\vu(\tau)}_{L^2}
    +2^{k\hn}\norm{\dk(\lam h\dive\vu)(\tau)}_{L^2}\Big)d\tau.
\end{align*}
Since $h\in L^\infty(\be[\hn+1])$, $\vu\in L^1(\be[\hn+1])$ and
$\lam h\dive\vu\in L^1(\be[\hn])$, we can get
\begin{align*}
    \sum_{k\in\Z}2^{k(\hn+1)}\norm{\dk h(t)}_{L^2}
    \lesssim &\norm{\rho_0-\rhobar}_{\be[\hn]}+(1+\norm{h}_{L^\infty(\be[\hn+1])})
    \norm{\vu}_{L^1(\be[\hn+1])}\\
    &+\norm{\lam h\dive\vu}_{L^1(\be[\hn])}<\infty.
\end{align*}
Thus, $\sum_{\abs{k}\ls N}\dk h$ converges uniformly in
$L^\infty(\Real^+; \be[\hn+1])$ and we can conclude that
$h\in\C(\Real^+;\be[\hn+1])$.

\subsection{Uniqueness}
Let $(h_1,\vu_1)$ and $(h_2,\vu_2)$ be solutions of
\begin{align}\label{uniq.1}
\left\{\begin{aligned}
    &h_t+\lam^{-1}(\vu\cdot\nabla \lam h )
        +\rhobar\lam^{-1}\dive\vu=-\lam^{-1}(\lam h\dive\vu),\\
    &\vu_t+\vu\cdot\nabla\vu-\rhobar^{-1}\mu\Delta\vu
    -\rhobar^{-1}(\mu+\lambda)\nabla\dive\vu
    +\lam\nabla h+\lam^{-1}\nabla h\\
    &\qquad\qquad\qquad\qquad
     =-\frac{\lam h}{\rhobar (\lam h+\rhobar)}
      (\mu\Delta \vu+(\mu+\lambda)\nabla\dive \vu).
\end{aligned}\right.
\end{align}
in $E_T^{\hn}$ with the same data
$(\lam^{-1}(\rho_0-\rhobar),\vu_0)$ constructed in the previous
parts on the time interval $[0,T]$. Denote $(\delta
h,\delta\vu)=(h_2-h_1,\vu_2-\vu_1)$. From the \eqref{uniq.1}, we can
get
\begin{align}\label{uniq.2}
\left\{\begin{aligned}
    &\partial_t\delta h+\lam^{-1}(\vu_2\cdot\nabla\lam h_2)
        +\rhobar\lam^{-1}\dive\delta\vu=F_1,\\
    &\partial_t\delta\vu+\vu_2\cdot\nabla\delta\vu
        +\rhobar^{-1} \mu\Delta \delta\vu
        -\rhobar^{-1}(\mu+\lambda)\nabla\dive\delta\vu\\
        &\qquad\qquad\qquad\qquad\qquad\qquad+\lam \nabla\delta h+\lam^{-1}\nabla\delta h=F_2,\\
    &(\delta h,\delta\vu)=(0,\mathbf{0}),
\end{aligned}\right.
\end{align}
where
\begin{align*}
    F_1=&-\lam^{-1}(\delta\vu\cdot\nabla h_1)
         -\lam^{-1}(\lam\delta h\dive\vu_2)-\lam^{-1}
          (\lam h_1\dive\delta\vu),\\
    F_2=&-\delta\vu\cdot\nabla\vu_1
      -\frac{\lam h_1}{\rhobar(\lam h_2+\rhobar)}
       (\mu\Delta\delta\vu+(\mu+\lambda)\nabla\dive \delta\vu)\\
    &+\left(\frac{1}{\lam h_2+\rhobar}
     -\frac{1}{\rhobar}-\frac{1}{\lam h_1+\rhobar}
     +\frac{1}{\rhobar}\right)
      (\mu\Delta\vu_1+(\mu+\lambda)\nabla\dive\vu_1).
\end{align*}

Similar to \eqref{nsp1}, we can get
\begin{align*}
    &\norm{(\delta h,\delta\vu)}_{E_T^\hn}\\
    \ls
    & Ce^{C\norm{\vu_2}_{L_T^1(\be[\hn+1])}}
     \left(\norm{F_1}_{L_T^1(\hbe[\hn-\frac{3}{2}]{\hn+1})}
    +\norm{F_2}_{L_T^1(\hbe[\hn-\frac{3}{2}]{\hn-1})}\right).
\end{align*}
Noticing that
\begin{align*}
  &h_1,\,h_2\in L_T^\infty(\hbe[\hn-\frac{3}{2}]{\hn+1})
   \cap L_T^1(\hbe[\hn+\frac{1}{2}]{\hn+1}),\\
  &\vu_1,\, \vu_2\in L_T^\infty(\hbe[\hn-\frac{3}{2}]{\hn-1})
   \cap L_T^1(\hbe[\hn+\frac{1}{2}]{\hn+1}),
\end{align*}
and
\begin{align*}
    \norm{h_1}_{L^\infty([0,T]\times\Real^N)}\ls \frac{1}{2}\rhobar,\quad
    \norm{h_2}_{L^\infty([0,T]\times\Real^N)}\ls \frac{1}{2}\rhobar,
\end{align*}
by the construction of solutions,
 we have with the help of interpolation arguments
\begin{align*}
    \norm{F_1}_{L_T^1(\hbe[\hn-\frac{3}{2}]{\hn+1})}
    \lesssim& \norm{h_1}_{L_T^2(\hbe[\hn-\frac{1}{2}]{\hn+1})}
    \norm{\delta\vu}_{L_T^2(\hbe[\hn-\frac{1}{2}]{\hn})}\\
    &+\norm{\delta h}_{L_T^\infty(\hbe[\hn-\frac{1}{2}]{\hn+1})}
    \norm{\vu_2}_{L_T^1(\hbe[\hn+\frac{1}{2}]{\hn+1})}\\
    &+\norm{h_1}_{L_T^\infty(\hbe[\hn-\frac{1}{2}]{\hn+1})}
    \norm{\delta\vu}_{L_T^1(\hbe[\hn+\frac{1}{2}]{\hn+1})},
\end{align*}
and
\begin{align*}
    &\norm{F_2}_{L_T^1(\hbe[\hn-\frac{3}{2}]{\hn-1})}\\
    \lesssim&\norm{\delta\vu}_{L_T^\infty(\hbe[\hn-\frac{3}{2}]{\hn-1})}
    \norm{\vu_1}_{L_T^1(\be[\hn+1])}
    +(1+\norm{h_2}_{L_T^\infty(\be[\hn+1])})\\
    &\cdot\norm{h_1}_{L_T^\infty(\be[\hn+1])}
    \norm{\delta\vu}_{L_T^1(\hbe[\hn+\frac{1}{2}]{\hn+1})}\\
    &+(\norm{h_1}_{L_T^\infty(\be[\hn+1])}
    +\norm{h_2}_{L_T^\infty(\be[\hn+1])})\norm{\delta h}_{L_T^\infty(\be[\hn+1])}
    \norm{\vu_1}_{L_T^1(\hbe[\hn+\frac{1}{2}]{\hn+1})}.
\end{align*}
Thus, we obtain
\begin{align*}
     \norm{(\delta h,\delta\vu)}_{E_T^\hn}\ls
     Ce^{C\norm{\vu_2}_{L_T^1(\be[\hn+1])}}
     \Big\{&(1+\norm{h_2}_{L_T^\infty(\be[\hn+1])})\\
     &\cdot \norm{h_1}_{L_T^\infty(\be[\hn+1])}+Z(T)\Big\}
     \norm{(\delta h,\delta\vu)}_{E_T^{\hn}},
\end{align*}
where $\limsup_{T\to 0^+} Z(T)=0$.

Supposing that $2C(1+\rhobar C_1^{-1})A\te[C]E(0)<\frac{1}{4}$
besides the conditions in \eqref{uni.8}, and taking $T>0$ small
enough such that $C\norm{\vu_2}_{L_T^1(\be[\hn+1])}\ls\ln 2$ and
$Z(T)<\frac{1}{2}$, we obtain $\norm{(\delta
h,\delta\vu)}_{E_T^\hn}\equiv 0$. Hence, $(h_1,\vu_1)\equiv
(h_2,\vu_2)$ on $[0,T]$.

Let $T_m$ (supposedly finite) be the largest time such that the two
solutions coincide on $[0,T_m]$. If we denote
\begin{align*}
    (\tilde{h}_i(t),
    \tilde{\vu}_i(t)):=(h_i(t+T_m),\vu_i(t+T_m)), \quad i=1,2,
\end{align*}
we can use the above arguments and the fact that
\begin{align*}
    \norm{\tilde{h}_i}_{L^\infty(\Real^+\times\Real^N)}\ls
\frac{1}{2}\rhobar \quad \text{ and}\quad \norm{\tilde{h}_i}_{L^\infty(\Real^+;
\hbe[\hn-\frac{3}{2}]{\hn+1})} \ls A\te[C]E(0)
\end{align*}
to prove that
$(\tilde{h}_1,\tilde{\vu}_1)=(\tilde{h}_2,\tilde{\vu}_2)$ on the
interval $[0,T_m]$ with the same $T_m$ as in the previous.
Therefore, we complete the proofs.

\section*{Acknowledgments}
The authors would like to thank the referees for their helpful
comments and suggestions on the manuscript.

C.C.Hao was partially supported by  the National Natural Science
Foundation of China (grants no. 10601061 and 10871134), the
Scientific Research Startup Special Foundation for the Winner of the
Award for Excellent Doctoral Dissertation and the Prize of President
Scholarship of Chinese Academy of Sciences (CAS) and the Fields
Frontier Project for Talented Youth of CAS. H.-L.Li was partially
supported by the National Natural Science Foundation of China
(grants no. 10431060 and 10871134), the Beijing Nova program,  the
NCET support of the Ministry of Education of China,  and the Huo
Ying Dong Foundation 111033.

\end{document}